\documentclass[conference]{IEEEtran}

\usepackage[letterpaper, left=0.625in, right=0.625in, bottom=1.04in, top=0.73in, columnsep=0.26in]{geometry}

\usepackage{url}
\usepackage[utf8]{inputenc}
\usepackage{xcolor}
\usepackage{amsmath}

\usepackage{multirow}
\usepackage{amssymb}
\usepackage{cite}
\usepackage{enumitem}
\usepackage{bm}
\usepackage{graphicx}
\usepackage{amsthm}

\usepackage{gensymb}

\usepackage[acronyms,nonumberlist,nopostdot,nomain,nogroupskip]{glossaries}
\usepackage{tablefootnote}
\usepackage{booktabs}
\usepackage{tabularx}
\usepackage{epsfig}
\usepackage{tikz}
\usetikzlibrary{external}
\usepackage{pgfplots}
\pgfplotsset{compat=newest} 
\pgfplotsset{plot coordinates/math parser=false}

\usetikzlibrary{plotmarks,patterns,patterns.meta,decorations.pathreplacing,backgrounds,calc,arrows,arrows.meta,spy,matrix}
\usepgfplotslibrary{patchplots,groupplots}
\usepackage{tikzscale}
\usepackage{siunitx}
\usepackage{tablefootnote}

\usepackage{multirow}
\usepackage{algorithm2e, setspace}
\SetAlCapNameFnt{\footnotesize}
\SetAlCapFnt{\footnotesize}
\RestyleAlgo{ruled}
\SetKwComment{Comment}{/* }{ */}
\usepackage{algpseudocode}

\usepackage[font=scriptsize]{subcaption}
\usepackage[font=footnotesize]{caption}

\usepackage{mathtools}

\usepackage{dblfloatfix}    
\usepackage{colortbl}
\usepackage{flushend}
\usepackage{footnote}

\usepackage{lipsum,afterpage,refcount}

\newacronym{3gpp}{3GPP}{3rd Generation Partnership Project}
\newacronym{adc}{ADC}{Analog to Digital Converter}
\newacronym{5g}{5G}{5th generation}
\newacronym{6g}{6G}{6th generation}
\newacronym{ai}{AI}{Artificial Intelligence}
\newacronym{aimd}{AIMD}{Additive Increase Multiplicative Decrease}
\newacronym{am}{AM}{Acknowledged Mode}
\newacronym{amc}{AMC}{Adaptive Modulation and Coding}
\newacronym{aqm}{AQM}{Active Queue Management}
\newacronym{awgn}{AGWN}{Additive White Gaussian Noise}
\newacronym{balia}{BALIA}{Balanced Link Adaptation}
\newacronym{bdp}{BDP}{Bandwidth-Delay Product}
\newacronym{bf}{BF}{beamforming}
\newacronym{cc}{CC}{Congestion Control}
\newacronym{cdf}{CDF}{Cumulative Distribution Function}
\newacronym{cn}{CN}{Core Network}
\newacronym{cqi}{CQI}{Channel Quality Information}
\newacronym{cp}{CP}{Control Plane}
\newacronym{csirs}{CSI-RS}{Channel State Information - Reference Signal}
\newacronym{dc}{DC}{Dual Connectivity}
\newacronym{rb}{RB}{Resource Block}
\newacronym{dce}{DCE}{Direct Code Execution}
\newacronym{dci}{DCI}{Downlink Control Information}
\newacronym{udp}{UDP}{User Datagram Protocol}
\newacronym{dl}{DL}{downlink}
\newacronym{fcfs}{FCFS}{first-come-first-served}
\newacronym{dmr}{DMR}{Deadline Miss Ratio}
\newacronym{fspl}{FSPL}{free-space path loss}
\newacronym{dmrs}{DMRS}{DeModulation Reference Signal}
\newacronym{e2e}{E2E}{End-to-End}
\newacronym{ppp}{PPP}{Poission Point Process}
\newacronym{aoi}{AoI}{Area of Interest}
\newacronym{cpu}{CPU}{Central Processing Unit}
 \newacronym{gpu}{GPU}{Graphics Processing Unit}
 \newacronym{tpu}{TPU}{Tensor Processing Unit}
\newacronym{si}{SI}{Study Item}
\newacronym{ecn}{ECN}{Explicit Congestion Notification}
\newacronym{edf}{EDF}{Earliest Deadline First}
\newacronym{enb}{eNB}{eNodeB}
\newacronym{epc}{EPC}{Evolved Packet Core}
\newacronym{es}{ES}{Edge Server}
\newacronym{cav}{CAV}{Connected and Autonomous Vehicle}
\newacronym{fdma}{FDMA}{Frequency Division Multiple Access}
\newacronym{fdd}{FDD}{Frequency Division Duplexing}
\newacronym{upa}{UPA}{Uniform Planar Array}
\newacronym{car}{CAR}{Circular Aperture Reflector }
\newacronym[firstplural=Radio Access Technologies (RATs)]{rat}{RAT}{Radio Access Technology}
\newacronym[firstplural=Radio Access Technology (RTs)]{rt}{RT}{Radio Technology}
\newacronym{fs}{FS}{Fast Switching}
\newacronym{isd}{ISD}{inter-site distance}
\newacronym{ftp}{FTP}{File Transfer Protocol}
\newacronym{gnb}{gNB}{Next Generation Node Base}
\newacronym{harq}{HARQ}{Hybrid Automatic Repeat reQuest}
\newacronym{hetnet}{HetNet}{Heterogeneous Network}
\newacronym{hh}{HH}{Hard Handover}
\newacronym{hol}{HOL}{Head-of-Line}
\newacronym{ia}{IA}{Initial Access}
\newacronym{imt}{IMT}{International Mobile Telecommunication}
\newacronym{iot}{IoT}{Internet of Things}
\newacronym{los}{LOS}{Line of Sight}
\newacronym{lte}{LTE}{Long Term Evolution}
\newacronym{m2m}{M2M}{Machine to Machine}
\newacronym{mac}{MAC}{Medium Access Control}
\newacronym{mc}{MC}{Multi-Connectivity}
\newacronym{mcs}{MCS}{Modulation and Coding Scheme}
\newacronym{mec}{MEC}{Mobile Edge Cloud}
\newacronym{mi}{MI}{Mutual Information}
\newacronym{mimo}{MIMO}{Multiple Input Multiple Output}
\newacronym{mmwave}{mmWave}{millimeter wave}
\newacronym{mptcp}{MPTCP}{Multipath TCP}
\newacronym{mr}{MR}{Maximum Rate}
\newacronym{mss}{MSS}{Maximum Segment Size}
\newacronym{mtd}{MTD}{Machine-Type Device}
\newacronym{mtu}{MTU}{Maximum Transmission Unit}
\newacronym{nfv}{NFV}{Network Function Virtualization}
\newacronym{vnf}{VNF}{Virtualization Network Function}
\newacronym{gv}{GV}{ground vehicle}
\newacronym{gvs}{GVs}{ground vehicles}
\newacronym{vec}{VEC}{Vehicular Edge Computing}
\newacronym{sdn}{SDN}{Software Defined Networking}
\newacronym{nlos}{NLOS}{Non Line of Sight}
\newacronym{nlosb}{NLOSb}{Building Non Line of Sight}
\newacronym{nlosv}{NLOSv}{Vehicle Non Line of Sight}
\newacronym{nr}{NR}{New Radio}
\newacronym{ofdm}{OFDM}{Orthogonal Frequency Division Multiplexing}
\newacronym{pdcch}{PDCCH}{Physical Downlonk Control Channel}
\newacronym{pdcp}{PDCP}{Packet Data Convergence Protocol}
\newacronym{pdsch}{PDSCH}{Physical Downlink Shared Channel}
\newacronym{pdu}{PDU}{Packet Data Unit}
\newacronym{pf}{PF}{Proportional Fair}
\newacronym{pgw}{PGW}{Packet Gateway}
\newacronym{phy}{PHY}{Physical}
\newacronym{pbch}{PBCH}{Physical Broadcast Channel}
\newacronym[plural=\gls{mme}s,firstplural=Mobility Management Entities (MMEs)]{mme}{MME}{Mobility Management Entity}
\newacronym{prb}{PRB}{Physical Resource Block}
\newacronym{pss}{PSS}{Primary Synchronization Signal}
\newacronym{pucch}{PUCCH}{Physical Uplink Control Channel}
\newacronym{pusch}{PUSCH}{Physical Uplink Shared Channel}
\newacronym{rach}{RACH}{Random Access Channel}
\newacronym{ran}{RAN}{Radio Access Network}
\newacronym{red}{RED}{Random Early Detection}
\newacronym{rf}{RF}{Radio Frequency}
\newacronym{rlc}{RLC}{Radio Link Control}
\newacronym{rlf}{RLF}{Radio Link Failure}
\newacronym{rrc}{RRC}{Radio Resource Control}
\newacronym{rrm}{RRM}{Radio Resource Management}
\newacronym{rr}{RR}{Round Robin}
\newacronym{rs}{RS}{Remote Server}
\newacronym{rsrp}{RSRP}{Reference Signal Received Power}
\newacronym{rss}{RSS}{Received Signal Strength}
\newacronym{rtt}{RTT}{Round Trip Time}
\newacronym{rw}{RW}{Receive Window}
\newacronym{rx}{RX}{Receiver}
\newacronym{sa}{SA}{standalone}
\newacronym{sack}{SACK}{Selective Acknowledgment}
\newacronym{sap}{SAP}{Service Access Point}
\newacronym{sch}{SCH}{Secondary Cell Handover}
\newacronym{scoot}{SCOOT}{Split Cycle Offset Optimization Technique}
\newacronym{sdma}{SDMA}{Spatial Division Multiple Access}
\newacronym{sinr}{SINR}{Signal to Interference plus Noise Ratio}
\newacronym{sm}{SM}{Saturation Mode}
\newacronym{snr}{SNR}{Signal-to-Noise Ratio}
\newacronym{son}{SON}{Self-Organizing Network}
\newacronym{ss}{SS}{Synchronization Signal}
\newacronym{srs}{SRS}{Sounding Reference Signal}
\newacronym{sss}{SSS}{Secondary Synchronization Signal}
\newacronym{tb}{TB}{Transport Block}
\newacronym{tcp}{TCP}{Transmission Control Protocol}
\newacronym{tdd}{TDD}{Time Division Duplexing}
\newacronym{tdma}{TDMA}{Time Division Multiple Access}
\newacronym{tfl}{TfL}{Transport for London}
\newacronym{tm}{TM}{Transparent Mode}
\newacronym{prr}{PRR}{Packet Reception Ratio}
\newacronym{trp}{TRP}{Transmitter Receiver Pair}
\newacronym{tti}{TTI}{Transmission Time Interval}
\newacronym{ttt}{TTT}{Time-to-Trigger}
\newacronym{tx}{TX}{Transmitter}
\newacronym{ue}{UE}{User Equipment}
\newacronym{ul}{UL}{uplink}
\newacronym{uml}{UML}{Unified Modeling Language}
\newacronym{um}{UM}{Unacknowledged Mode}
\newacronym{utc}{UTC}{Urban Traffic Control}
\newacronym{vm}{VM}{Virtual Machine}
\newacronym{rsrq}{RSRQ}{Reference Signal Received Quality}
\newacronym{rssi}{RSSI}{Received Signal Strength Indicator}
\newacronym{crs}{CRS}{Cell Reference Signal}
\newacronym{v2v}{V2V}{Vehicle-to-Vehicle}
\newacronym{v2i}{V2I}{Vehicle-to-Infrastructure}
\newacronym{v2n}{V2N}{Vehicle-to-Network}
\newacronym{v2x}{V2X}{Vehicle-to-Everything}
\newacronym{vn}{VN}{Vehicular Node}
\newacronym{dsrc}{DSRC}{Dedicated Short Range Communication}
\newacronym{ci}{CI}{context information}
\newacronym{voi}{VoI}{value of information}
\newacronym{gps}{GPS}{Global Positioning System}
\newacronym{qos}{QoS}{Quality of Service}
\newacronym{qoe}{QoE}{Quality of Experience}
\newacronym{ml}{ML}{Machine Learning}
\newacronym{ahp}{AHP}{Analytic Hierarchy Process}
\newacronym{lidar}{LIDAR}{Light Detection and Ranging}
\newacronym{sumo}{SUMO}{Simulation of Urban MObility}
\newacronym{wave}{WAVE}{Wireless Access in Vehicular Environment}
\newacronym{c-its}{C-ITS}{Connected Intelligent Transportation System}
\newacronym{dash}{DASH}{Dynamic Adaptive Streaming over HTTP}
\newacronym{http}{HTTP}{HyperText Transfer Protocol}
\newacronym{nt}{NT}{Non-Terrestrial}
\newacronym{ntc}{NTC}{non-terrestrial communication}
\newacronym{ntn}{NTN}{Non-Terrestrial Network}
\newacronym{hap}{HAP}{High Altitude Platform}
\newacronym{leo}{LEO}{Low Earth Orbit}
\newacronym{meo}{MEO}{Medium Earth Orbit}
\newacronym{geo}{GEO}{Geostationary Earth Orbit}
\newacronym{uav}{UAV}{Unmanned Aerial Vehicle}
\newacronym{nsat}{nSAT}{Nanosatellite}
\newacronym{ehf}{EHF}{extremely high-frequency}
\newacronym{ioe}{IoE}{Internet of Everyone}
\newacronym{gan}{GaN}{Gallium Nitride}
\newacronym{tle}{TLE}{two-line element}
\newacronym{ecdf}{ECDF}{Empirical Cumulative Distribution Function}
\newacronym{fifo}{FIFO}{First-Input First-Output}

\usepackage{tikz}
\usepackage{pgfplots}
\usepackage{tikzscale}
\usepackage{tikz-qtree}

\pgfplotsset{compat=newest}
\pgfplotsset{plot coordinates/math parser=false}
\pgfplotsset{every axis/.append style={
                    label style={font=\scriptsize},
                    tick label style={font=\scriptsize},
                    legend style={font=\scriptsize}
                    }}
\usetikzlibrary{plotmarks,shapes,patterns,decorations.pathreplacing,backgrounds,calc,arrows,arrows.meta,spy,matrix,shadows,trees,positioning,fit}
\usepgfplotslibrary{patchplots,groupplots}

\tikzstyle{startstop} = [rectangle, rounded corners, minimum width=2cm, minimum height=0.5cm,text centered, draw=black]
\tikzstyle{io} = [trapezium, trapezium left angle=70, trapezium right angle=110, minimum width=3cm, minimum height=1cm, text centered, draw=black]
\tikzstyle{process} = [rectangle, minimum width=2cm, minimum height=0.5cm, text centered, draw=black, alignb=center]
\tikzstyle{decision} = [ellipse, minimum width=2cm, minimum height=1cm, text centered, draw=black]
\tikzstyle{arrow} = [thick,<->,>=stealth]
\tikzstyle{line} = [thick,>=stealth]
\tikzstyle{darrow} = [thick,<->,>=stealth,dashed]
\tikzstyle{sarrow} = [thick,->,>=stealth]
\tikzstyle{larrow} = [line width=0.1mm,dashdotted,->,>=stealth]

\pgfkeys{/pgf/number format/.cd,1000 sep={\,}} 

\makeatletter
\def\grd@save@target#1{%
  \def\grd@target{#1}}
\def\grd@save@start#1{%
  \def\grd@start{#1}}
\tikzset{
  grid with coordinates/.style={
    to path={%
      \pgfextra{%
        \edef\grd@@target{(\tikztotarget)}%
        \tikz@scan@one@point\grd@save@target\grd@@target\relax
        \edef\grd@@start{(\tikztostart)}%
        \tikz@scan@one@point\grd@save@start\grd@@start\relax
        \draw[minor help lines] (\tikztostart) grid (\tikztotarget);
        \draw[major help lines] (\tikztostart) grid (\tikztotarget);
        \grd@start
        \pgfmathsetmacro{\grd@xa}{\the\pgf@x/1cm}
        \pgfmathsetmacro{\grd@ya}{\the\pgf@y/1cm}
        \grd@target
        \pgfmathsetmacro{\grd@xb}{\the\pgf@x/1cm}
        \pgfmathsetmacro{\grd@yb}{\the\pgf@y/1cm}
        \pgfmathsetmacro{\grd@xc}{\grd@xa + \pgfkeysvalueof{/tikz/grid with coordinates/major step x}}
        \pgfmathsetmacro{\grd@yc}{\grd@ya + \pgfkeysvalueof{/tikz/grid with coordinates/major step y}}
        \foreach \x in {\grd@xa,\grd@xc,...,\grd@xb}
        \node[anchor=north] at (\x,\grd@ya) {\pgfmathprintnumber{\x}};
        \foreach \y in {\grd@ya,\grd@yc,...,\grd@yb}
        \node[anchor=east] at (\grd@xa,\y) {\pgfmathprintnumber{\y}};
      }
    }
  },
  minor help lines/.style={
    help lines,
    gray,
    line cap =round,
    xstep=\pgfkeysvalueof{/tikz/grid with coordinates/minor step x},
    ystep=\pgfkeysvalueof{/tikz/grid with coordinates/minor step y}
  },
  major help lines/.style={
    help lines,
    line cap =round,
    line width=\pgfkeysvalueof{/tikz/grid with coordinates/major line width},
    xstep=\pgfkeysvalueof{/tikz/grid with coordinates/major step x},
    ystep=\pgfkeysvalueof{/tikz/grid with coordinates/major step y}
  },
  grid with coordinates/.cd,
  minor step x/.initial=.5,
  minor step y/.initial=.2,
  major step x/.initial=1,
  major step y/.initial=1,
  major line width/.initial=1pt,
}
\makeatother

\newlength\fheight
\newlength\fwidth

\definecolor{steelblue}{RGB}{176,196,222}

\usepackage[capitalize]{cleveref}
\crefname{section}{Sec.}{Secs.}
\usetikzlibrary{decorations}

\IEEEoverridecommandlockouts

\newcommand\copyrightnotice{%
\begin{tikzpicture}[remember picture,overlay]
\node[anchor=south,yshift=15pt] at (current page.south) {\fbox{\parbox{\dimexpr\textwidth-\fboxsep-\fboxrule\relax}{
\footnotesize \textcopyright 2025 IEEE. Personal use of this material is permitted.
Permission from IEEE must be obtained for all other uses, in any current or future media,
including reprinting/republishing this material for advertising or promotional purposes,
creating new collective works, for resale or redistribution to servers or lists,
or reuse of any copyrighted component of this work in other works.}}};
\end{tikzpicture}
}

\begin{document}
\bstctlcite{IEEEexample:BSTcontrol}

\title{Performance Evaluation of Satellite-Based \\ Data Offloading on Starlink Constellations\vspace{-0.3cm}}

    \author{\IEEEauthorblockN{Alexander Bonora, Alessandro Traspadini, Marco Giordani, Michele Zorzi\medskip}

\IEEEauthorblockA{ Department of Information Engineering, University of Padova, Italy.\\
Email:	\texttt{\{bonora, traspadini, giordani, zorzi\}@dei.unipd.it\vspace{-0.4cm}}
\thanks{This work was supported by the European Commission through the European Union’s Horizon Europe Research and Innovation Programme under the Marie Skłodowska-Curie-SE, Grant Agreement No. 101129618, UNITE. This research was also partially supported by the European Union under the Italian National Recovery and Resilience Plan (NRRP) of NextGenerationEU, 
partnership on ``Telecommunications of the Future'' (PE0000001 - program “RESTART”).}}}

\maketitle

\copyrightnotice

\begin{abstract}
      \Gls{vec} is a key research area in autonomous driving. As Intelligent Transportation Systems (ITSs) continue to expand, \glspl{gv} face the challenge of handling huge amounts of sensor data to drive safely. Specifically, due to energy and capacity limitations, \glspl{gv} will need to offload resource-hungry tasks to external (cloud) computing units for faster processing.
      In \gls{6g} wireless systems, the research community is exploring the concept of \glspl{ntn}, where satellites can serve as space edge computing nodes to aggregate, store, and process data from \glspl{gv}. 
      In this paper we propose new data offloading strategies between a cluster of \glspl{gv} and satellites in the \glspl{leo}, to optimize the trade-off between coverage and end-to-end delay. For the accuracy of the simulations, we consider real data and orbits from the Starlink constellation, one of the most representative and popular examples of commercial satellite deployments for communication. 
      Our results demonstrate that Starlink satellites can support real-time offloading under certain conditions that depend on the onboard computational capacity of the satellites, the frame rate of the sensors, and the number of GVs.

\end{abstract}

\glsresetall


\begin{tikzpicture}[remember picture,overlay]
\node[anchor=north,yshift=-10pt] at (current page.north) {\parbox{\dimexpr\textwidth-\fboxsep-\fboxrule\relax}{
\centering\footnotesize 
This paper has been accepted for publication in the 2025 IEEE Wireless Communications and Networking Conference (WCNC). \textcopyright 2025 IEEE.\\
Please cite it as: A. Bonora, A. Traspadini, M. Giordani, and M. Zorzi, "Performance Evaluation of Satellite-Based Data Offloading on Starlink Constellations," in Proc. IEEE Wireless Communications and Networking Conference (WCNC), 2025.}};
\end{tikzpicture}

\glsresetall

\section{Introduction}
\label{sec:intro}

One primary goal of \gls{6g} networks is to ensure global broadband Internet access~\cite{giordani2020toward}, which is challenging today due to several technological, economic, and geographical reasons~\cite{Chaoub20216g}.
From a technological point of view, current networks mainly rely on terrestrial infrastructure, which may be difficult to deploy in remote areas, e.g., oceans or deserts.
Additionally, harsh weather and terrain in some rural regions, e.g., in the countryside, as well as the lack of efficient power grids and electricity, further complicate tower installation.
From an economic perspective, network deployment in unserved regions is expensive, and the return on investment for operators and/or service providers is not guaranteed. 
One possible solution for 6G is to leverage \glspl{ntn}~\cite{giordani2021non}, where aerial and space nodes such as \glspl{uav}, \glspl{hap}, and satellites provide global Internet access from the sky. For example, \glspl{ntn} can promote on-demand connectivity where terrestrial infrastructure is unavailable, and complement terrestrial networks in case of emergency~\cite{wang2020potential}.

In particular, \gls{leo} satellites are an attractive option for \glspl{ntn}. Compared to \gls{geo} satellites, LEOs operate in closer proximity to the Earth, at an altitude between 300 and 1\,000 km, which ensures lower latency (typically around 30 to 50 ms), higher data throughput, and better signal quality. At the same time, LEO satellites can shape very large coverage umbrellas on the ground (of several hundreds of kilometers of radius), which is crucial to provide global coverage. As evidence of this, there exist today many commercial Internet access deployments based on LEO satellites, including Starlink
 by SpaceX or OneWeb
 by Eutelsat. With more than 3 million customers as of May 2024, and with residential subscription plans as competitive as $120$ USD per month in the US, Starlink is one of the most representative and popular examples of satellite-based Internet solutions, with more than $5\,000$ satellites already in orbit (with a plan to have up to $42\,000$ by 2030)~\cite{li2023networking,michel2022first}. 

Besides providing connectivity, \glspl{ntn} can act as edge servers for processing, caching, and/or storing data generated from power- and capacity-constrained ground devices~\cite{9509294}.
In particular, teleoperated driving~\cite{ZHANG20164} relies on the availability of massive data from sensors onboard the \glspl{gv} to guarantee accurate perception of the environment~\cite{aoki2020cooperative}. However, data processing based on machine learning (from compression and object detection and recognition to tracking and trajectory prediction) requires extensive computational resources, which may be challenging for \glspl{gv}~\cite{testolina2023selma}.
In an urban scenario, \glspl{gv} can offload data to roadside units for \gls{vec}~\cite{liu2021vehicular} but, in poorly connected rural areas, \glspl{ntn} emerge as a viable alternative~\cite{traspadini2022uavhapassisted}.
To address this research, in~\cite{Traspadini23Real} we proposed a framework to optimize data offloading via \glspl{ntn}, focusing on \glspl{hap}. This framework accounts for latency and computational capacity constraints, and aims at maximizing the probability of processing data in real-time. Similarly, Qiao \emph{et al.}~\cite{Qi18Colla} proposed a collaborative framework to offload computationally-intensive tasks to heterogeneous \gls{vec} platforms to guarantee low communication and computation latency. Soret \emph{et al.} also proposed to use \gls{leo} satellites for data offloading and backhauling, and the problem was addressed based on an Age of Information approach~\cite{Soret215G}.
In~\cite{Orbit22Cassara}, the authors presented a task offloading algorithm in mega-\gls{leo} satellite constellations to enhance task distribution and efficiency.
To date, these papers make strict assumptions on the channel model, constellation topology, and orbit dynamics, which may give misleading performance~results.

The main contributions of this paper are twofold.
First, we propose a novel \gls{vec} framework for \glspl{gv} that includes mechanisms to dynamically offload computing tasks to \gls{leo} satellites if it improves real-time communication and processing delay, or prioritize onboard processing otherwise. This framework also incorporates dropping and back-off policies to periodically decongest satellite servers and ensure efficient resource utilization.
Second, we conduct a realistic evaluation using real orbital traces and parameters from the Starlink constellation, coupled with the implementation of the 3GPP TR 38.811 channel model.
Our simulation results validate the feasibility of using Starlink satellites as edge computing servers, demonstrate their ability to accelerate data processing, and provide practical guidelines for system dimensioning in terms of computational capacity, density of the satellite constellation, and application frame rate.

In detail, Sec.~\ref{sec:system_model} presents our system model, Sec.~\ref{sec:offloading} describes our VEC offloading strategy, Sec.~\ref{sec:performance_evaluation} discusses the simulation results, and Sec.~\ref{sec:conclusions_and_future_works} summarizes our conclusions.
\section{System Model}
\label{sec:system_model}

\begin{figure}[t!]
\centering 
\includegraphics[width=0.97\columnwidth]{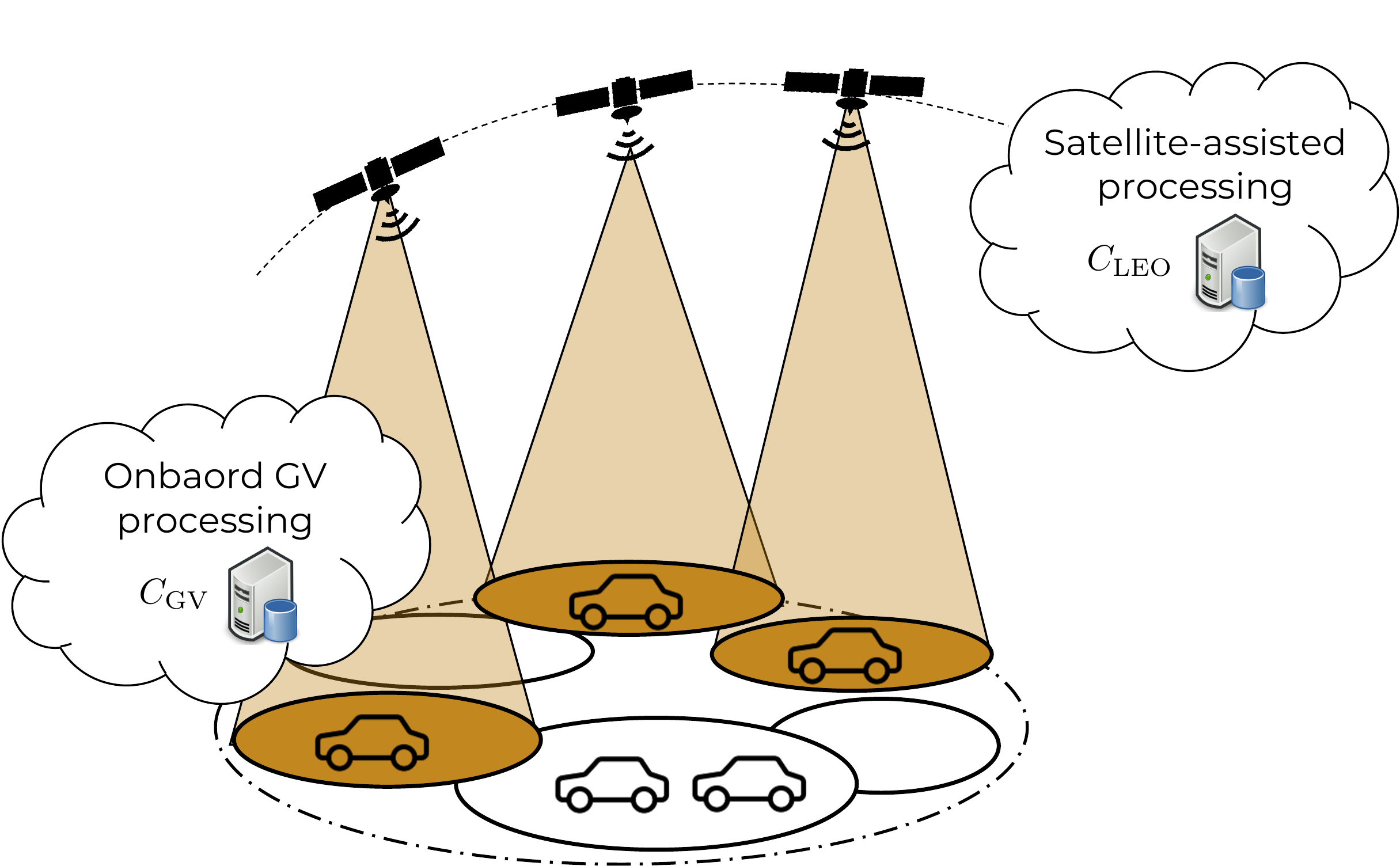}
\caption{Illustration of the scenario. We deploy $n$ GVs in a rural/unserved area under the coverage of a constellation of Starlink LEO satellites. GVs (satellites) are equipped with a computing platform with capacity $C_{\rm GV}$ ($C_{\rm LEO}$) for processing data.}
\label{fig:scenario}
\vspace{-1.5em}
\end{figure}

This section describes the research problem and scenario (\cref{sub:scenario}), the satellite orbit model (\cref{sub:orbit}), the channel model (\cref{sub:channel}), and the delay model (\cref{sub:delay}).

\subsection{Scenario Description}
\label{sub:scenario}
Our scenario, depicted in Fig.~\ref{fig:scenario}, consists of a set of $n$ \glspl{gv} that collect data through onboard sensors at a fixed inter-frame rate $r$.
Each frame involves a constant computational load $C$ (e.g., for object detection), and must be processed within the latency constraints of the application $\delta$.
Each frame can be processed onboard the \gls{gv}, or offloaded to a \gls{leo} Starlink satellite.
On the one hand, data offloading can reduce the processing delay as \gls{leo} satellites are less constrained, and can reasonably mount more powerful computing platforms (with capacity $C_{\rm LEO}$) compared to \glspl{gv} (with capacity $C_{\rm GV}$), so we have $C_{\rm LEO}\gg C_{\rm GV}$. On the other hand, it also introduces a non-negligible delay for both data offloading from the \gls{gv} to the remote server at the LEO, and the return of the processed data back to the \gls{gv}.

We consider a constellation of $s$ \gls{leo} Starlink satellites, and each \gls{gv} can carefully select a satellite for data offloading among those that are in coverage.
As described later in~\cref{sec:offloading}, based on the state of the local queues and the delays measured on previously processed data from the LEO satellites, each \gls{gv} can determine the optimal offloading policy to maximize the probability that data is processed within the latency constraint~$\delta$.
\subsection{Orbit Model and Deployment}
\label{sub:orbit}
The position of a \gls{gv} is determined by its latitude $y_v$ and longitude $x_v$, while the satellite's position is defined by its altitude $h_s$, along with the latitude $y_s$ and the longitude $x_s$ of its projection on the surface of the Earth.
The position of a Starlink satellite on its orbit can be precisely derived based on the \gls{tle} data from \url{celestrak.org}, which provides real-time and up-to-date satellite trajectories from Starlink orbital parameters, to calculate the position and velocity of the nodes in space~\cite{Hong16TLE}.

The distance $d$ between a \gls{gv} and a generic satellite can be computed as:
\begin{equation}
    d = (h_s + R_E) \sqrt{1+\left(\frac{R_E}{h_s + R_E}\right)^2 -\frac{2 R_E}{h_s + R_E} \cos(\alpha)},
\end{equation}
where $R_E$ is the Earth's radius (i.e., 6\,371~km), and $\alpha$ represents the angle between the \gls{gv} and the satellite as observed from the Earth's center. The cosine of $\alpha$ can be derived as:
\begin{equation}
    \cos(\alpha) = \cos(y_s) \cos(y_v) \cos(x_v - x_s) + \sin(y_v) \sin(y_s).
\end{equation}
From~\cite{geyer13Geometric}, the elevation angle $\theta$ can be calculated as:
\begin{equation}
    \theta = \arccos\left(\frac{(h_s + R_E) \sin(\alpha)}{d}\right)
\end{equation}

\subsection{Channel Model}

\label{sub:channel}
According to the 3GPP specifications~\cite{38821}, ground-to-satellite connectivity can be established in the high-capacity \gls{mmwave} bands, thereby enabling multi-Gbps data rates~\cite{giordani2020satellite}.
Assuming \gls{los}, the \gls{snr} in dB between a transmitter $i$ and a receiver $j$ is given by
\begin{equation}
\gamma_{i,j} = \text{EIRP}_{i} + (G_j/T) - \text{PL}_{i,j} - k - B,
\label{eq:snr}
\end{equation}
where $\text{EIRP}_{i}$ is the effective isotropic radiated power of the transmitter in \SI{}{\watt}, $(G_j/T)$ is the receive antenna-gain-to-noise-temperature, $\text{PL}_{i,j}$ is the path loss, $k$ is the Boltzmann constant and $B$ is the bandwidth in \SI{}{\hertz} (all quantities are on a log scale).
For the ground-to-satellite channel, the path loss includes several attenuation components,  especially the scintillation loss ($\text{PL}_s$) due to changes in the refractive index in the atmosphere, and the atmospheric absorption loss ($\text{PL}_g$) due to dry air and water vapor attenuation. Overall, the path loss is~\cite{38821,3GPP_38811}
\begin{equation}
\text{PL} = \text{FSPL} + \text{PL}_s + \text{PL}_g,
\end{equation}
where $\text{FSPL}$ is the free space path loss given by
\begin{equation}
    \text{FSPL} = 92.45 + 20\log(f_c)+ 20 \log(d),
\end{equation}
$d$ is the distance in km, and $f_c$ is the carrier frequency in GHz.
From early results in~\cite{giordani2020satellite}, we proved that the several stages of attenuation introduced by the atmosphere at high frequency and over long distance, such as in ground-to-satellite channels, can be mitigated by highly directional antennas and the resulting beamforming gain to improve the link quality.

In the case of offloading, from the \gls{snr} in~\cref{eq:snr} we can derive the transmission delays $t_{\rm UL}$ and $t_{\rm DL}$ for each data frame in uplink (i.e., from the GV to the satellite) and downlink (i.e., from the satellite to the GV), respectively. We have that $t_{k} = n_k / R_k$, $k\in{\rm UL, \rm DL}$, where $n_k$ is the size of the transmitted data, and $R_k$ is the ergodic capacity given by $B\log_2(1+\gamma_{i,j})$.
\subsection{Delay Model}
\label{sub:delay}
Both the LEO satellites and the \glspl{gv} are equipped with infinitely long \gls{fifo} queues with deterministic service time equal to $C/C_{\rm LEO}$ and $C/C_{\rm GV}$, respectively.
In the case of onboard processing, no transmission is involved. 
Therefore, the total delay $t_d$ can be written as
\begin{equation}
t_d = t_{\rm GV} = W^{\rm GV}_q + {C}/{C_{\rm GV}},
\end{equation}
where $W^{\rm GV}_q$ is the queuing delay, while $C/C_{\rm GV}$ is the onboard processing time of a single frame. 

In the case of offloading, the total delay depends on many components, and we have that
\begin{equation}
t_d = t_{\rm LEO} =2\tau_p + t_{\rm UL} + t_{\rm DL} + W^{\rm LEO}_q + {C}/{C_{\rm LEO}},
\end{equation}
where $W^{\rm LEO}_q$ and $C/C_{\rm LEO}$ are the queuing delay and the processing delay at the satellite, $\tau_p$ is the propagation delay, and $t_{\rm UL}$ and $t_{\rm DL}$ are the UL and DL transmission delays as described in Sec.~\ref{sub:channel}.

From $t_d$, we can evaluate the real-time probability $P_{RT}$, defined as the probability that the total delay is within the time constraint $\delta$ specified by the application, i.e.,
\begin{equation}
	P_{RT} = \mathbb{P}(t_d < \delta).
\end{equation}
\section{Satellite Offloading Strategies}
\label{sec:offloading}

Data processing involves three options: onboard processing, offloading to a \gls{leo} satellite, or dropping. 
In general, onboard processing is more convenient, provided that sufficient computational resources are available at the \glspl{gv}, so as to avoid the additional delay and potential communication overhead associated with transmitting data to and from remote servers. 
Indeed, data offloading is triggered only if onboard processing is infeasible within the time constraint \(\delta\) of the application.
Given the local occupancy of its queue, each GV can estimate $\hat{W}^{\rm GV}_q$ and so the onboard delay \(\hat{t}_{\rm GV}\):
if \(\hat{t}_{\rm GV} < \delta\), data are processed onboard, otherwise are either offloaded or dropped.

In the case of offloading, the GV directly connects to a LEO satellite according to either of the following methods:
\begin{itemize}
\item Maximum-\gls{snr} ({MS}): Each \gls{gv} selects the satellite with the highest \gls{snr} $\gamma$, where $\gamma$ was defined in Eq.~\eqref{eq:snr}.
\item Sufficient-Random ({SR}): Each \gls{gv} selects a random satellite in visibility, i.e., among those whose \gls{snr} is above a pre-defined threshold $\gamma^s_{th}$, which guarantees a more distributed selection of the satellites.
\end{itemize}
In both methods, each \gls{gv} maintains connectivity with the selected satellite until it goes out of coverage, i.e., as long as $\gamma > \gamma_{th}$, where $\gamma_{th}$ depends on the sensitivity of the receiver.
Given the highly dynamic nature of the \gls{leo} networks, we utilize a feedback mechanism to predict the current load at the satellite~\cite{LEOSAN2022}.
In this sense, proper data scheduling and back-off strategies are crucial for promoting low latency during offloading.
We propose and evaluate two strategies.

\paragraph{Back-Off Offloading (BOO)}
A feedback mechanism is implemented between each \gls{gv} and its serving \gls{leo} satellite to monitor the evolution of the system.
The feedback incorporates the state/occupancy of the queue at the satellite, so the \gls{gv} can estimate the queuing waiting time $\hat{W}^{\rm LEO}_q$, which is inversely proportional to the available buffer capacity. With the assumption that the delay of the link is known a priori, as the distance $d$ to the serving satellite is also known, the GV can estimate the total delay for data offloading~\(\hat{t}_{\rm LEO}\).

If $\hat{t}_{\rm LEO} < \delta$, data are offloaded to the \gls{leo} satellite.
On the contrary, if $\hat{t}_{\rm LEO} \geq \delta$, the \gls{leo} satellite is overloaded, thus data offloading is not possible. As such, data will be dropped. 
Notably, BOO has been designed so that data offloading is deactivated for $t_o$ frames, in order to sufficiently reduce the burden at the satellite, and permit the queue to decongest.
In this scenario with strict delay constraints, we claim that it is more convenient to discard data that cannot be delivered on time (and therefore would no longer be relevant or useful for the application), rather than consuming transmission and processing resources unnecessarily.
The value of $t_o$ is given in terms of number of frames, and is uniformly distributed within a pre-defined interval, i.e.,
\begin{equation}
    t_o \thicksim U(1, t^{m}_o),
\end{equation}
where $t^{m}_o$ is a system parameter. 
If no feedback is received within $\delta$, the most recent feedback is considered obsolete, and the \gls{gv} also enters a back-off condition where communication with the satellite is deactivated for the following $t_o$ frames.

In addition, if the most recent feedback is older than $\delta$ (i.e., the feedback is obsolete) and the vehicle is not in back-off condition, the \gls{gv} cannot determine the current state of the satellite queue.
Consequently, for the next frame transmission, it estimates $\hat{t}_{\rm LEO}$ under the assumption that $\hat{W}^{\rm LEO}_q=0$. Once the feedback related to this packet is received, the \gls{gv} updates its estimate of $\hat{W}^{\rm LEO}_q$.

\paragraph{Light-Drop and Back-Off Offloading (LDBOO)} 
In addition to the back-off mechanism proposed in BOO, LDBOO also implements a dropping policy even when $\hat{t}_{\rm LEO} < \delta$.
This is to prevent the satellites' queues from overloading.
In this scheme, data are dropped with probability
\begin{equation}
{p}_{\rm drop} =\left({\hat{t}_{\rm LEO}}/{\delta}\right)^\sigma,
\label{eq:drop}
\end{equation}
where $\sigma$ is a parameter that describes the steepness of ${p}_{\rm drop}$.
Thus, as $\hat{t}_{\rm LEO}$ approaches $\delta$, i.e., as the system becomes progressively more congested, ${p}_{\rm drop}$ increases, thereby alleviating the processing load at the satellite.

Notice that BOO drops data only when the system is already congested. Moreover, BOO only relies on feedback notifications that take at least $\tau_p$ to propagate and, therefore, do not describe the current state of the queue.\footnote{Notice that, in the satellite scenario, $\tau_p$ may be several tens of ms, so notifications may be severely obsolete.}
On the contrary, LDBOO mitigates congestion by dropping data even before the satellite queues overflow, which could improve the delay.

In terms of overhead, both BOO and LDBOO require the satellite to estimate the state of the local queues, although this calculation is typically less resource-intensive compared to the actual processing tasks. Moreover, the choice of the vehicle to either drop or process a packet introduces a slight communication overhead, which is necessary for the vehicle to correctly assess the available resources, even though this decision step is negligible compared to the transmission time.

\section{Performance Evaluation}

\label{sec:performance_evaluation}
In Sec.~\ref{sub:sim-params} we describe our simulation parameters, while performance results are presented in Sec.~\ref{sec:simulation_results}.

\subsection{Simulation Setup and Parameters}
\label{sub:sim-params}

\begin{table}[t!]
  \centering
  \caption{Simulation parameters.}
  \label{tab:parameters}
  \small
  \renewcommand{\arraystretch}{1.1}
  \begin{tabular}{|l|l|}
    \hline
    {Parameter}                              & {Value}                 \\ \hline
    Packet size ($n_{\rm UL}$) [Mb]                           & 3                            \\
    Packet size ($n_{\rm DL}$) [Mb]                          & 0.1                          \\
    Computational load ($C$) [TFLOPS]                             & 0.06                           \\
    Computational capacity ($C_{\rm GV}$) [TFLOPS]           & 0.5                              \\
    Computational capacity ($C_{\rm LEO}$) [TFLOPS]              & \{5, 10 15, 20\}             \\
    Application time constraint ($\delta$) [s]                                       & 0.15                          \\
    Satellite constellation size ($s$)                        & \{2831, 5662\}  \\
    Number of \glspl{gv} ($n$)                            & \{10, 50, 100\}                     \\
    Sensor frame rate ($r$) [fps]                           & \{10, 30\}                       \\
    Simulation time [s]                      & 60 \\\hline
    Satellite antenna ($G_j/T$) [dB/K]      & 15.84                  \\
    \gls{gv} antenna ($G_j/T$) [dB/K]        & 19.19                \\
    Carrier frequency ($f_c$) [GHz]                          & 30                        \\
    Bandwidth ($B$) [MHz]                                   & 10                        \\
    EIRP satellite antenna [dBW]                          & 34.9                       \\
    EIRP \gls{gv} antenna [dBW]                            & 37.2                       \\ 
    Earth radius ($R_E$) [km]                            & 6371                        \\ 
    Satellite height ($h_s$) [km]                            &  350$-$600                       \\ 
    SNR SR policy threshold  ($\gamma^s_{th}$) [dB]                            &  10                       \\ 
    SNR threshold ($\gamma_{th}$) [dB]                            &  0                       \\ \hline  
  \end{tabular}
  \vspace{-1.5em}
\end{table}
\label{sec:simulation_results}
Simulation parameters, if not specified otherwise, are reported in~\cref{tab:parameters}.
Each GV produces a sensor's frame (e.g., an RGB camera image) of size $n_{\rm UL}$ = 3~Mb, at rates ${r} = 10$ or $30$~fps. This value of $n_{\rm UL}$ is compatible with real-world vehicular data: for example, according to the SELMA dataset~\cite{testolina2023selma}, the size of a raw RGB camera frame is approximately 20 Mb, which reduces to around 3~Mb after compression.
Each frame involves a constant computational load ${C} = 60$ GFLOP (giga floating point operations) for processing (e.g., for object detection and classification), and the processed output (e.g., bounding boxes) is eventually returned to the \glspl{gv} in a packet of size $n_{\rm DL}= 0.1$~Mb $\ll n_{\rm UL}$. The computational capacity of the GV is $C_{\rm GV}=0.5$~TFLOPS, while for the LEO satellite we tested different configurations varying from  $C_{\rm LEO}=5$ to $20$~TFLOPS, which is consistent with the previous literature on this topic~\cite{Traspadini23Real}. Additionally, the application delay constraint is fixed to $\delta=0.15$~s.\footnote{While the average human reaction time to visual stimuli is approximately $0.2$-$0.25$ s, the optimal reaction time for teleoperated driving software depends on several factors. 
For instance, at 60 km/h, a delay of $\delta=0.15$~s results in the vehicle traveling about 2.5 meters before taking an action, which is sufficient for the vehicle to stop safely in case of an emergency.}
GVs communicate at \glspl{mmwave} at a frequency of $f_c=38$~GHz through orthogonal subcarriers of width $B=10$~MHz.

Simulation results are given as a function of the size of the Starlink satellite constellation, for $s=5662$ (i.e., 100\% coverage) and $s=2831$ (i.e., 50\% coverage) satellites. 
The height of the satellite $h_s$ is in the range  350$-$600 km, depending on the considered constellation. Furthermore, we investigate the impact of the number of \glspl{gv} $n$ in the scenario, the frame rate $r$, the satellite selection policy (MS vs. SR) and the offloading policy (BOO vs. LDBOO). We also evaluate the effect of different values of $\sigma$ and $t^m_o$.

\subsection{Simulation Results}
\label{sec:simulation_results}

\begin{figure}[t!]
\centering
  \setlength\fwidth{\columnwidth}
\begin{tikzpicture}

\definecolor{crimson2143940}{RGB}{214,39,40}
\definecolor{darkgray176}{RGB}{176,176,176}
\definecolor{darkorange25512714}{RGB}{255,127,14}
\definecolor{forestgreen4416044}{RGB}{44,160,44}
\definecolor{mediumpurple148103189}{RGB}{148,103,189}
\definecolor{orchid227119194}{RGB}{227,119,194}
\definecolor{sienna1408675}{RGB}{140,86,75}
\definecolor{steelblue31119180}{RGB}{31,119,180}

\begin{axis}[
width = \columnwidth,
height = 4.3cm,
tick align=outside,
tick pos=left,
x grid style={darkgray176},
xlabel={Time [s]},
xmajorgrids,
xmin=0, xmax=60,
xtick style={color=black},
y grid style={darkgray176},
ylabel={Elevation angle ($\theta$) [deg]},
ymajorgrids,
ymin=40.4498180327869, ymax=87.0278213114754,
ytick={40,50,60,70,80,90},
ytick style={color=black},
legend style={legend cell align=left,
              align=center,
              draw=white!15!black,
              at={(0.5, 1.3)},
              anchor=center,
              /tikz/every even column/.append style={column sep=1em}},
legend columns=2,
]
\addplot [line width=1.2pt, steelblue31119180]
table {%
0 42.567
1 42.6654086021505
2 42.8736648044693
3 43.080455026455
4 43.2863076923077
5 43.4736596858639
6 43.6805721925134
7 43.8668839779005
8 44.0672544378698
9 44.2405663265306
10 44.4245
11 44.5928640776699
12 44.7721508379888
13 44.9439411764706
14 45.1235348837209
15 45.2689764705882
16 45.4361818181818
17 45.5920883977901
18 45.7412058823529
19 45.8857967914439
20 46.028238317757
21 46.1671313131313
22 46.2960909090909
23 46.4145769230769
24 46.5417487684729
25 46.64835
26 46.7577464788732
27 46.8584055299539
28 46.9627336244542
29 47.0529469026549
30 47.1334084507042
31 47.2127741935484
32 47.2845
33 47.3448955223881
34 47.4058695652174
35 47.4680277777778
36 47.5121463414634
37 47.553
38 47.5866382978723
39 47.6145
40 47.6354595959596
41 47.6498229166667
42 47.6574926829268
43 47.658
44 47.651417721519
45 47.6355969581749
46 47.6181516245487
47 47.5926123188406
48 47.5589665427509
49 47.5220788530466
50 47.4762357414449
51 47.4226785714286
52 47.369520754717
53 47.3190724637681
};
\addlegendentry{STARLINK-2607}
\addplot [line width=1.2pt, darkorange25512714]
table {%
0 51.5901313868613
1 51.7461916167665
2 51.91625
3 52.0753260869565
4 52.2172121212121
5 52.3601690821256
6 52.5028480392157
7 52.6374285714286
8 52.7588304093567
9 52.8739090909091
10 52.9687722222222
11 53.0771082802548
12 53.158
13 53.2349788359788
14 53.3035047169811
15 53.3655757575758
16 53.4066666666667
17 53.4444444444444
18 53.4762264150943
19 53.4971284916201
20 53.5110869565217
21 53.5085966850829
22 53.4970120481928
23 53.4795674157303
24 53.4533168316832
25 53.4108666666667
26 53.3587802197802
27 53.2977610619469
28 53.2325090909091
29 53.1582431192661
30 53.0729473684211
31 52.9785789473684
32 52.8795921787709
33 52.7702264150943
34 52.6411506849315
35 52.5169261083744
36 52.3738591549296
37 52.2490560747664
38 52.0661421568627
39 51.9128846153846
40 51.7501176470588
41 51.575786407767
42 51.3841162790698
43 51.1902110091743
44 51.0036603773585
45 50.7927372881356
46 50.6013783783784
47 50.3820373831776
48 50.1637735849057
49 49.9489054726368
50 49.7154516129032
51 49.599
};
\addlegendentry{STARLINK-3624}
\addplot [line width=1.2pt, forestgreen4416044]
table {%
0 57.2189285714286
1 56.8561551724138
2 56.3895530726257
3 55.8774976076555
4 55.4550372340426
5 54.979118556701
6 54.5409192825112
7 54.0526972477064
8 53.6165721925134
9 53.1435639534884
10 52.6800769230769
11 52.2378674698795
12 51.8018670520231
13 51.3280561797753
14 50.9147662337662
15 50.4476242038217
16 50.0355
17 49.62
18 49.199078313253
19 48.7147142857143
20 48.3241428571429
21 47.9042931034483
22 47.4573487179487
23 47.0662527472528
24 46.650112244898
25 46.2541847826087
26 46.051
};
\addlegendentry{STARLINK-3625}
\addplot [line width=1.2pt, crimson2143940]
table {%
0 72.0228424242424
1 71.820417721519
2 71.5290939226519
3 71.225912568306
4 70.935627027027
5 70.5899395348837
6 70.244602739726
7 69.8713461538461
8 69.4955879396985
9 69.132875
10 68.6995635359116
11 68.2620967741936
12 67.835
13 67.40546875
14 66.9163146067416
15 66.496
16 66.0137566137566
17 65.5034117647059
18 65.0333404255319
19 64.6124035087719
20 64.1113555555555
21 63.60864
22 63.1222588235294
23 62.655676300578
24 62.1251794871795
25 61.6521202185792
26 61.0705652173913
27 60.6299186602871
28 60.1610786026201
29 59.6756744186047
30 59.1565
31 58.6517050691244
32 58.1444418604651
33 57.6632081447964
34 57.1574864864865
35 56.6745769230769
36 56.1877512437811
37 55.6895962441315
38 55.2577351598174
39 54.7514558823529
40 54.2619863013699
41 53.7623130434783
42 53.341783919598
43 52.8691428571429
44 52.4012804878049
45 51.9148609865471
46 51.443311627907
47 51.0216119402985
48 50.5699074074074
49 50.07699
50 49.7016728971963
51 49.2056842105263
52 48.8028674242424
53 48.3781338289963
54 47.9193092369478
55 47.5091535580524
56 47.0700681818182
57 46.680562962963
58 46.2774431818182
59 45.8481918819188
60 45.522888
};
\addlegendentry{STARLINK-3818}
\addplot [line width=1.2pt, sienna1408675]
table {%
0 84.9106393442623
1 84.3502122905028
2 83.6810467836257
3 83.0285675675676
4 82.3545043859649
5 81.7110746268657
6 81.0084597156398
7 80.3057260273973
8 79.6438314606742
9 78.9211666666667
10 78.214493902439
11 77.6023563218391
12 76.8098048780488
13 76.2000650887574
14 75.501705882353
15 74.7971602209945
16 74.1470877192982
17 73.4330726256983
18 72.7955112359551
19 72.1684747474747
20 71.4629447513812
21 70.8531204819277
22 70.1872954545454
23 69.541371257485
24 68.8635964912281
25 68.2498875739645
26 67.5953606557377
27 66.9991651376147
28 66.3506422018349
29 65.75577
30 65.1124019607843
31 64.4944702970297
32 63.9153652173913
33 63.2987623318386
34 62.6754038461538
35 62.1040936170213
36 61.5676513761468
37 60.9361809954751
38 60.3876651162791
39 59.8424784688995
40 59.2697192982456
41 58.6924215686275
42 58.1558547008547
43 57.6347818181818
44 57.0939907407407
45 56.5513693693694
46 56.0061566820277
47 55.4943842592593
48 54.9475174129353
49 54.4421009615385
50 53.9172631578947
51 53.4266192468619
52 52.9572472727273
53 52.42695
54 51.9637258687259
55 51.4758897058824
56 51.0204208494209
57 50.5466065573771
58 50.0651318681319
59 49.6077564575646
60 49.2683768115942
};
\addlegendentry{STARLINK-5328}
\addplot [line width=1.2pt, orchid227119194]
table {%
0 67.1971230769231
1 67.6539371428571
2 68.1915586592179
3 68.7534157894737
4 69.2804940239044
5 69.7923130841122
6 70.3297444444444
7 70.8242189054726
8 71.3492245989305
9 71.8620304568528
10 72.3590946745562
11 72.7858538011696
12 73.2672099447514
13 73.7669016393443
14 74.1862905027933
15 74.6031040462428
16 75.0129940828402
17 75.3898012048193
18 75.7722095808383
19 76.1091428571429
20 76.3990975609756
21 76.6723604651163
22 76.9325
23 77.1399657142857
24 77.3233146853147
25 77.4511117318436
26 77.5535394736842
27 77.6155
28 77.6324347826087
29 77.6024075829384
30 77.5389661835749
31 77.4235229357798
32 77.286847161572
33 77.096972972973
34 76.8823813953488
35 76.6432964824121
36 76.3455145631068
37 76.0249473684211
38 75.6935841584158
39 75.2962093023256
40 74.9188648648649
41 74.5583888888889
42 74.0685477386935
43 73.6563532338308
44 73.1711717171717
45 72.7288285714286
46 72.2133393665158
47 71.7309444444444
48 71.1919567099567
49 70.7032966507177
50 70.1838847926267
51 69.6161357466063
52 69.1448576388889
53 68.5666720647773
54 68.051
55 67.511572519084
56 66.9604349442379
57 66.4109117647059
58 65.8664981684982
59 65.3183798449612
60 64.8872919708029
};
\addlegendentry{STARLINK-6342}
\end{axis}

\end{tikzpicture}
    \vspace{-1.5em}
    \label{fig:elev_time}
    \caption{Elevation angle over time for some Starlink satellites (with IDs).\vspace{-0.5em}}
\end{figure}
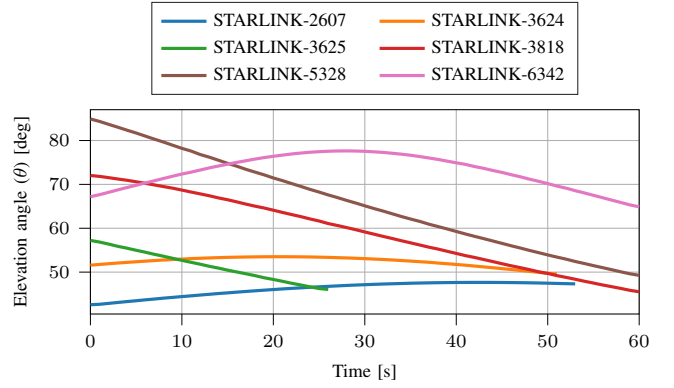

\begin{figure}[t!]
    \centering
  {
\begin{tikzpicture}
\pgfplotsset{every tick label/.append style={font=\scriptsize}}

\pgfplotsset{compat=1.11,
  /pgfplots/ybar legend/.style={
    /pgfplots/legend image code/.code={%
      \draw[##1,/tikz/.cd,yshift=-0.5em]
      (0cm,0cm) rectangle (10pt,0.6em);},
  },
}

\definecolor{burlywood246183156}{RGB}{246,183,156}
\definecolor{coral23011690}{RGB}{230,116,90}
\definecolor{cornflowerblue111145242}{RGB}{111,145,242}
\definecolor{darkcyan32144140}{RGB}{32,144,140}
\definecolor{darkgray176}{RGB}{176,176,176}
\definecolor{darkslateblue48103141}{RGB}{48,103,141}
\definecolor{darkslateblue6857130}{RGB}{68,57,130}
\definecolor{gainsboro221220219}{RGB}{221,220,219}
\definecolor{lightblue170198253}{RGB}{170,198,253}
\definecolor{lightgray204}{RGB}{204,204,204}
\definecolor{mediumseagreen53183120}{RGB}{53,183,120}
\definecolor{yellowgreen14421467}{RGB}{144,214,67}
\definecolor{color1}{RGB}{21, 52, 69}
\definecolor{color2}{RGB}{78, 161, 207}
\definecolor{color3}{RGB}{23,108,155}
\definecolor{color4}{RGB}{158, 211, 240}
\definecolor{color6giallo}{RGB}{194,135,32}

\begin{axis}[
width = \columnwidth/1.15,
height = 5cm,
legend style={legend cell align=left,
              align=center,
              draw=white!15!black,
              at={(0.5, 1.3)},
              anchor=center,
              /tikz/every even column/.append style={column sep=1em}},
legend columns=3,
tick align=outside,
tick pos=left,
x grid style={darkgray176},
ymajorgrids,
xlabel={Maximum timeout length (\(\displaystyle t^m_o\))},
xmin=-0.59, xmax=3.59,
xtick style={color=black},
xtick={0,1,2,3},
xticklabels={1,5,10,30},
y grid style={darkgray176},
ylabel={Real-time probability ($P_{\rm RT}$) [\%]},
ymin=0, ymax=110,
ytick style={color=black}
]
\draw[draw=black,fill=coral23011690] (axis cs:0.24,0) rectangle (axis cs:0.4,77.9975213675214);
\draw[draw=black,fill=coral23011690] (axis cs:1.24,0) rectangle (axis cs:1.4,84.3846581196581);
\draw[draw=black,fill=coral23011690] (axis cs:2.24,0) rectangle (axis cs:2.4,85.2059829059829);
\draw[draw=black,fill=coral23011690] (axis cs:3.24,0) rectangle (axis cs:3.4,84.2829487179487);
\draw[draw=black,fill=color1,fill opacity=0.6] (axis cs:0.08,0) rectangle (axis cs:0.24,84.5618376068376);
\draw[draw=black,fill=color1,fill opacity=0.6] (axis cs:1.08,0) rectangle (axis cs:1.24,87.5867094017094);
\draw[draw=black,fill=color1,fill opacity=0.6] (axis cs:2.08,0) rectangle (axis cs:2.24,87.4611965811966);
\draw[draw=black,fill=color1,fill opacity=0.6] (axis cs:3.08,0) rectangle (axis cs:3.24,86.9547435897436);
\draw[draw=black,fill=color3] (axis cs:-0.08,0) rectangle (axis cs:0.08,85.04);
\draw[draw=black,fill=color3] (axis cs:0.92,0) rectangle (axis cs:1.08,86.9146581196581);
\draw[draw=black,fill=color3] (axis cs:1.92,0) rectangle (axis cs:2.08,87.6226495726496);
\draw[draw=black,fill=color3] (axis cs:2.92,0) rectangle (axis cs:3.08,87.0532478632479);
\draw[draw=black,fill=color2] (axis cs:-0.24,0) rectangle (axis cs:-0.08,80.4220085470085);
\draw[draw=black,fill=color2] (axis cs:0.76,0) rectangle (axis cs:0.92,80.6145299145299);
\draw[draw=black,fill=color2] (axis cs:1.76,0) rectangle (axis cs:1.92,81.2043162393162);
\draw[draw=black,fill=color2] (axis cs:2.76,0) rectangle (axis cs:2.92,81.2677777777778);
\draw[draw=black,fill=color4] (axis cs:-0.4,0) rectangle (axis cs:-0.24,68.2343162393162);
\draw[draw=black,fill=color4] (axis cs:0.6,0) rectangle (axis cs:0.76,68.1097008547008);
\draw[draw=black,fill=color4] (axis cs:1.6,0) rectangle (axis cs:1.76,68.2300427350427);
\draw[draw=black,fill=color4] (axis cs:2.6,0) rectangle (axis cs:2.76,68.1769230769231);

\addplot[ybar,ybar legend,draw=black,fill=color4,line width=0.08pt]
table[row sep=crcr]{%
  0 0\\
};
\addlegendentry{$\sigma=1$ (LDBOO)}
\addplot[ybar,ybar legend,draw=black,fill=color2,line width=0.08pt]
table[row sep=crcr]{%
  0 0\\
};
\addlegendentry{$\sigma=2$ (LDBOO)}
\addplot[ybar,ybar legend,draw=black,fill=color3,line width=0.08pt]
table[row sep=crcr]{%
  0 0\\
};
\addlegendentry{$\sigma=4$ (LDBOO)}
\addplot[ybar,ybar legend,draw=black,fill=color1,fill opacity=0.6,line width=0.08pt]
table[row sep=crcr]{%
  0 0\\
};
\addlegendentry{$\sigma=6$ (LDBOO)}

 \addplot[ybar,ybar legend,draw=black,fill=coral23011690,line width=0.08pt]
table[row sep=crcr]{%
   0 0\\
};
\addlegendentry{BOO}

\addplot[ybar,ybar legend,draw=black,fill=white,line width=0.08pt]
table[row sep=crcr]{%
  0 0\\
};
\addlegendentry{$P_{\rm RT}$}
\addplot[ybar,ybar legend,draw=black,fill=white,pattern=north east lines,line width=0.08pt]
table[row sep=crcr]{%
  0 0\\
};
\addlegendentry{$P_{\rm D}$}

\end{axis}

\begin{axis}[
width = \columnwidth/1.15,
height = 5cm,
legend cell align={left},
legend style={
  fill opacity=0.8,
  draw opacity=1,
  text opacity=1,
  at={(1.2,0.7)},
  anchor=north west,
  draw=lightgray204
},
tick align=outside,
x grid style={darkgray176},
xmin=-0.59, xmax=3.59,
xticklabel=\empty,
y grid style={darkgray176},
ylabel={\scriptsize Data drop probability ($P_{\rm D}$) [\%]},
ymin=0, ymax=110,
ytick pos=right,
ytick style={color=black},
yticklabel style={anchor=west}
]
\draw[draw=black,fill=yellowgreen14421467,pattern=north east lines,fill opacity=0.8] (axis cs:0.28,0) rectangle (axis cs:0.36,6.71200567863005);
\draw[draw=black,fill=yellowgreen14421467,pattern=north east lines,fill opacity=0.8] (axis cs:1.28,0) rectangle (axis cs:1.36,9.97311539610454);
\draw[draw=black,fill=yellowgreen14421467,pattern=north east lines,fill opacity=0.8] (axis cs:2.28,0) rectangle (axis cs:2.36,12.339182761033);
\draw[draw=black,fill=yellowgreen14421467,pattern=north east lines,fill opacity=0.8] (axis cs:3.28,0) rectangle (axis cs:3.36,15.6112326447724);
\draw[draw=black,fill=mediumseagreen53183120,pattern=north east lines,fill opacity=0.8] (axis cs:0.12,0) rectangle (axis cs:0.2,8.56354334891309);
\draw[draw=black,fill=mediumseagreen53183120,pattern=north east lines,fill opacity=0.8] (axis cs:1.12,0) rectangle (axis cs:1.2,10.0488418594693);
\draw[draw=black,fill=mediumseagreen53183120,pattern=north east lines,fill opacity=0.8] (axis cs:2.12,0) rectangle (axis cs:2.2,11.4125164903776);
\draw[draw=black,fill=mediumseagreen53183120,pattern=north east lines,fill opacity=0.8] (axis cs:3.12,0) rectangle (axis cs:3.2,13.0360628051014);
\draw[draw=black,fill=darkcyan32144140,pattern=north east lines,fill opacity=0.8] (axis cs:-0.04,0) rectangle (axis cs:0.04,10.0439700666795);
\draw[draw=black,fill=darkcyan32144140,pattern=north east lines,fill opacity=0.8] (axis cs:0.96,0) rectangle (axis cs:1.04,11.5807642817247);
\draw[draw=black,fill=darkcyan32144140,pattern=north east lines,fill opacity=0.8] (axis cs:1.96,0) rectangle (axis cs:2.04,12.1048238868274);
\draw[draw=black,fill=darkcyan32144140,pattern=north east lines,fill opacity=0.8] (axis cs:2.96,0) rectangle (axis cs:3.04,12.94670386893);
\draw[draw=black,fill=darkslateblue48103141,pattern=north east lines,fill opacity=0.8] (axis cs:-0.2,0) rectangle (axis cs:-0.12,18.2312742601392);
\draw[draw=black,fill=darkslateblue48103141,pattern=north east lines,fill opacity=0.8] (axis cs:0.8,0) rectangle (axis cs:0.88,18.4277699026624);
\draw[draw=black,fill=darkslateblue48103141,pattern=north east lines,fill opacity=0.8] (axis cs:1.8,0) rectangle (axis cs:1.88,18.5105476450651);
\draw[draw=black,fill=darkslateblue48103141,pattern=north east lines,fill opacity=0.8] (axis cs:2.8,0) rectangle (axis cs:2.88,18.7322142170025);
\draw[draw=black,fill=darkslateblue6857130,pattern=north east lines,fill opacity=0.8] (axis cs:-0.36,0) rectangle (axis cs:-0.28,31.765670185611);
\draw[draw=black,fill=darkslateblue6857130,pattern=north east lines,fill opacity=0.8] (axis cs:0.64,0) rectangle (axis cs:0.72,31.890285516972);
\draw[draw=black,fill=darkslateblue6857130,pattern=north east lines,fill opacity=0.8] (axis cs:1.64,0) rectangle (axis cs:1.72,31.7699436880583);
\draw[draw=black,fill=darkslateblue6857130,pattern=north east lines,fill opacity=0.8] (axis cs:2.64,0) rectangle (axis cs:2.72,31.8230633234772);
\end{axis}

\end{tikzpicture}
      \vspace{-0.6em}
      \label{fig:ecdf_1}
  }
    \vspace{-0.25cm} \caption{Real-time probability (solid bars) and data drop probability (striped bars) for different offloading strategies vs. $\sigma$ and $t^m_o$, with $r = 30$~fps, $C_{\rm LEO} = 20$~TFLOPS, and $n = 100$ GVs.\vspace{-0.5cm}}
   \label{fig:ecdf_plots}
\end{figure}
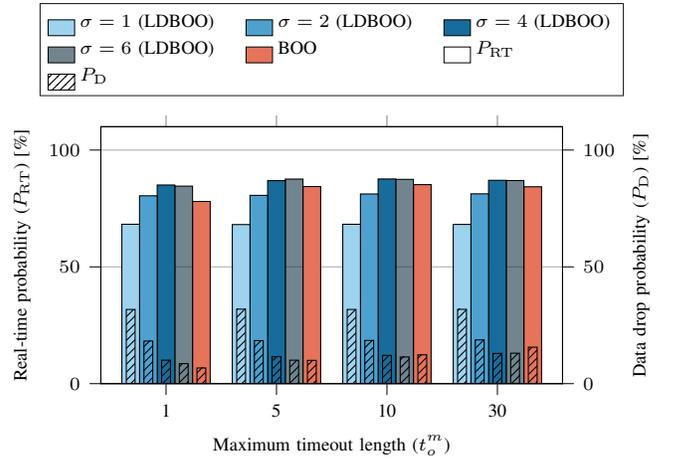

\begin{figure}[t!]
\hspace*{14pt}
	\begin{subfigure}[t]{\columnwidth}
		\centering
		\setlength\fwidth{\columnwidth}
%
%
\definecolor{color1}{RGB}{23,108,155}
\definecolor{color2}{RGB}{194,135,32}

\begin{tikzpicture}
\pgfplotsset{every tick label/.append style={font=\scriptsize}}

\pgfplotsset{compat=1.11,
	/pgfplots/ybar legend/.style={
		/pgfplots/legend image code/.code={%
			\draw[##1,/tikz/.cd,yshift=-0.25em]
			(0cm,0cm) rectangle (10pt,0.6em);},
	},
}

\begin{axis}[%
width=0,
height=0,
at={(0,0)},
scale only axis,
xmin=0,
xmax=0,
xtick={},
ymin=0,
ymax=0,
ytick={},
axis background/.style={fill=white},
legend style={legend cell align=left,
              align=center,
              draw=white!15!black,
              at={(1, 1.3)},
              anchor=center,
              /tikz/every even column/.append style={column sep=1em}},
legend columns=2,
]
\addplot[ybar,ybar legend,draw=black,fill=color1,line width=0.08pt]
table[row sep=crcr]{%
	0	0\\
};
\addlegendentry{$r=10$}

\addplot[ybar legend,ybar,draw=black,fill=color2,line width=0.08pt]
  table[row sep=crcr]{%
	0	0\\
};
\addlegendentry{$r=30$}

\end{axis}
\end{tikzpicture}%
	\end{subfigure}
	\centering
	\subfloat[][{SR} selection policy.]
	{
\begin{tikzpicture}

\definecolor{darkgoldenrod19413532}{RGB}{194,135,32}
\definecolor{darkgray176}{RGB}{176,176,176}
\definecolor{darkslategray54}{RGB}{54,54,54}
\definecolor{lightgray204}{RGB}{204,204,204}
\definecolor{teal23108155}{RGB}{23,108,155}

\begin{axis}[
width = \columnwidth,
height = 4.5cm,
legend cell align={left},
legend style={
  fill opacity=0.8,
  draw opacity=1,
  text opacity=1,
  at={(0.03,0.97)},
  anchor=north west,
  draw=lightgray204
},
tick align=outside,
tick pos=left,
x grid style={darkgray176},
xlabel={Number of \glspl{gv} ($n$)},
xmin=-0.5, xmax=2.5,
xtick style={color=black},
xtick={0,1,2},
xticklabels={10,50,100},
y grid style={darkgray176},
ymajorgrids,
ylabel={Delay ($t_d$) [s]},
ymin=-0.001, ymax=2.5,
ytick style={color=black},
name=ax1
]
\path [draw=darkslategray54, fill=teal23108155, semithick]
(axis cs:-0.396,0.0371)
--(axis cs:-0.004,0.0371)
--(axis cs:-0.004,0.0417)
--(axis cs:-0.396,0.0417)
--(axis cs:-0.396,0.0371)
--cycle;
\path [draw=darkslategray54, fill=darkgoldenrod19413532, semithick]
(axis cs:0.004,0.0374)
--(axis cs:0.396,0.0374)
--(axis cs:0.396,0.0424)
--(axis cs:0.004,0.0424)
--(axis cs:0.004,0.0374)
--cycle;
\path [draw=darkslategray54, fill=teal23108155, semithick]
(axis cs:0.604,0.0374)
--(axis cs:0.996,0.0374)
--(axis cs:0.996,0.0422)
--(axis cs:0.604,0.0422)
--(axis cs:0.604,0.0374)
--cycle;
\path [draw=darkslategray54, fill=darkgoldenrod19413532, semithick]
(axis cs:1.004,0.04)
--(axis cs:1.396,0.04)
--(axis cs:1.396,0.0545)
--(axis cs:1.004,0.0545)
--(axis cs:1.004,0.04)
--cycle;
\path [draw=darkslategray54, fill=teal23108155, semithick]
(axis cs:1.604,0.0377)
--(axis cs:1.996,0.0377)
--(axis cs:1.996,0.0432)
--(axis cs:1.604,0.0432)
--(axis cs:1.604,0.0377)
--cycle;
\path [draw=darkslategray54, fill=darkgoldenrod19413532, semithick]
(axis cs:2.004,0.045)
--(axis cs:2.396,0.045)
--(axis cs:2.396,0.9517)
--(axis cs:2.004,0.9517)
--(axis cs:2.004,0.045)
--cycle;
\draw[draw=darkslategray54,fill=teal23108155,line width=0.3pt] (axis cs:0,0) rectangle (axis cs:0,0);

\draw[draw=darkslategray54,fill=darkgoldenrod19413532,line width=0.3pt] (axis cs:0,0) rectangle (axis cs:0,0);

\addplot [semithick, darkslategray54, forget plot]
table {%
-0.2 0.0371
-0.2 0.0302
};
\addplot [semithick, darkslategray54, forget plot]
table {%
-0.2 0.0417
-0.2 0.0486
};
\addplot [semithick, darkslategray54, forget plot]
table {%
-0.298 0.0302
-0.102 0.0302
};
\addplot [semithick, darkslategray54, forget plot]
table {%
-0.298 0.0486
-0.102 0.0486
};
\addplot [semithick, darkslategray54, forget plot]
table {%
0.2 0.0374
0.2 0.03
};
\addplot [semithick, darkslategray54, forget plot]
table {%
0.2 0.0424
0.2 0.0499
};
\addplot [semithick, darkslategray54, forget plot]
table {%
0.102 0.03
0.298 0.03
};
\addplot [semithick, darkslategray54, forget plot]
table {%
0.102 0.0499
0.298 0.0499
};
\addplot [semithick, darkslategray54, forget plot]
table {%
0.8 0.0374
0.8 0.0303
};
\addplot [semithick, darkslategray54, forget plot]
table {%
0.8 0.0422
0.8 0.0494
};
\addplot [semithick, darkslategray54, forget plot]
table {%
0.702 0.0303
0.898 0.0303
};
\addplot [semithick, darkslategray54, forget plot]
table {%
0.702 0.0494
0.898 0.0494
};
\addplot [semithick, darkslategray54, forget plot]
table {%
1.2 0.04
1.2 0.0263
};
\addplot [semithick, darkslategray54, forget plot]
table {%
1.2 0.0545
1.2 0.0762
};
\addplot [semithick, darkslategray54, forget plot]
table {%
1.102 0.0263
1.298 0.0263
};
\addplot [semithick, darkslategray54, forget plot]
table {%
1.102 0.0762
1.298 0.0762
};
\addplot [semithick, darkslategray54, forget plot]
table {%
1.8 0.0377
1.8 0.0295
};
\addplot [semithick, darkslategray54, forget plot]
table {%
1.8 0.0432
1.8 0.0514
};
\addplot [semithick, darkslategray54, forget plot]
table {%
1.702 0.0295
1.898 0.0295
};
\addplot [semithick, darkslategray54, forget plot]
table {%
1.702 0.0514
1.898 0.0514
};
\addplot [semithick, darkslategray54, forget plot]
table {%
2.2 0.045
2.2 0.0263
};
\addplot [semithick, darkslategray54, forget plot]
table {%
2.2 0.9517
2.2 2.3117
};
\addplot [semithick, darkslategray54, forget plot]
table {%
2.102 0.0263
2.298 0.0263
};
\addplot [semithick, darkslategray54, forget plot]
table {%
2.102 2.3117
2.298 2.3117
};
\addplot [semithick, darkslategray54, forget plot]
table {%
-0.396 0.0393
-0.004 0.0393
};
\addplot [semithick, darkslategray54, forget plot]
table {%
0.004 0.0397
0.396 0.0397
};
\addplot [semithick, darkslategray54, forget plot]
table {%
0.604 0.0396
0.996 0.0396
};
\addplot [semithick, darkslategray54, forget plot]
table {%
1.004 0.0447
1.396 0.0447
};
\addplot [semithick, darkslategray54, forget plot]
table {%
1.604 0.0402
1.996 0.0402
};
\addplot [semithick, darkslategray54, forget plot]
table {%
2.004 0.0623
2.396 0.0623
};

\coordinate (c1) at (axis cs:-0.396,0);
\coordinate (c2) at (axis cs:2.396,0.1);
\draw (c1) rectangle (axis cs:2.396,0.1);

\end{axis}

\begin{axis}[
    width = \textwidth/3.7,
    height = 2.5cm,
   name=ax2,
   scaled ticks=false,
   xmin=-0.45,xmax=2.5,
   ymin=0,ymax=0.15,
   at={($(ax1.south west)+(1cm,1.5cm)$)},
   xmajorgrids=true,
   xtick={0, 1, 2},
   xticklabels={10, 50, 100},
   y grid style={darkgray176},
   ymajorgrids,
   ytick style={color=black},
   ytick={0, 0.05, 0.1},
   yticklabels={0, 0.05, 0.1},
   clip=true
 ]
\path [draw=darkslategray54, fill=teal23108155, semithick]
(axis cs:-0.396,0.0371)
--(axis cs:-0.004,0.0371)
--(axis cs:-0.004,0.0417)
--(axis cs:-0.396,0.0417)
--(axis cs:-0.396,0.0371)
--cycle;
\path [draw=darkslategray54, fill=darkgoldenrod19413532, semithick]
(axis cs:0.004,0.0374)
--(axis cs:0.396,0.0374)
--(axis cs:0.396,0.0424)
--(axis cs:0.004,0.0424)
--(axis cs:0.004,0.0374)
--cycle;
\path [draw=darkslategray54, fill=teal23108155, semithick]
(axis cs:0.604,0.0374)
--(axis cs:0.996,0.0374)
--(axis cs:0.996,0.0422)
--(axis cs:0.604,0.0422)
--(axis cs:0.604,0.0374)
--cycle;
\path [draw=darkslategray54, fill=darkgoldenrod19413532, semithick]
(axis cs:1.004,0.04)
--(axis cs:1.396,0.04)
--(axis cs:1.396,0.0545)
--(axis cs:1.004,0.0545)
--(axis cs:1.004,0.04)
--cycle;
\path [draw=darkslategray54, fill=teal23108155, semithick]
(axis cs:1.604,0.0377)
--(axis cs:1.996,0.0377)
--(axis cs:1.996,0.0432)
--(axis cs:1.604,0.0432)
--(axis cs:1.604,0.0377)
--cycle;
\path [draw=darkslategray54, fill=darkgoldenrod19413532, semithick]
(axis cs:2.004,0.045)
--(axis cs:2.396,0.045)
--(axis cs:2.396,0.9517)
--(axis cs:2.004,0.9517)
--(axis cs:2.004,0.045)
--cycle;
\draw[draw=darkslategray54,fill=teal23108155,line width=0.3pt] (axis cs:0,0) rectangle (axis cs:0,0);

\draw[draw=darkslategray54,fill=darkgoldenrod19413532,line width=0.3pt] (axis cs:0,0) rectangle (axis cs:0,0);

\addplot [semithick, darkslategray54, forget plot]
table {%
-0.2 0.0371
-0.2 0.0302
};
\addplot [semithick, darkslategray54, forget plot]
table {%
-0.2 0.0417
-0.2 0.0486
};
\addplot [semithick, darkslategray54, forget plot]
table {%
-0.298 0.0302
-0.102 0.0302
};
\addplot [semithick, darkslategray54, forget plot]
table {%
-0.298 0.0486
-0.102 0.0486
};
\addplot [semithick, darkslategray54, forget plot]
table {%
0.2 0.0374
0.2 0.03
};
\addplot [semithick, darkslategray54, forget plot]
table {%
0.2 0.0424
0.2 0.0499
};
\addplot [semithick, darkslategray54, forget plot]
table {%
0.102 0.03
0.298 0.03
};
\addplot [semithick, darkslategray54, forget plot]
table {%
0.102 0.0499
0.298 0.0499
};
\addplot [semithick, darkslategray54, forget plot]
table {%
0.8 0.0374
0.8 0.0303
};
\addplot [semithick, darkslategray54, forget plot]
table {%
0.8 0.0422
0.8 0.0494
};
\addplot [semithick, darkslategray54, forget plot]
table {%
0.702 0.0303
0.898 0.0303
};
\addplot [semithick, darkslategray54, forget plot]
table {%
0.702 0.0494
0.898 0.0494
};
\addplot [semithick, darkslategray54, forget plot]
table {%
1.2 0.04
1.2 0.0263
};
\addplot [semithick, darkslategray54, forget plot]
table {%
1.2 0.0545
1.2 0.0762
};
\addplot [semithick, darkslategray54, forget plot]
table {%
1.102 0.0263
1.298 0.0263
};
\addplot [semithick, darkslategray54, forget plot]
table {%
1.102 0.0762
1.298 0.0762
};
\addplot [semithick, darkslategray54, forget plot]
table {%
1.8 0.0377
1.8 0.0295
};
\addplot [semithick, darkslategray54, forget plot]
table {%
1.8 0.0432
1.8 0.0514
};
\addplot [semithick, darkslategray54, forget plot]
table {%
1.702 0.0295
1.898 0.0295
};
\addplot [semithick, darkslategray54, forget plot]
table {%
1.702 0.0514
1.898 0.0514
};
\addplot [semithick, darkslategray54, forget plot]
table {%
2.2 0.045
2.2 0.0263
};
\addplot [semithick, darkslategray54, forget plot]
table {%
2.2 0.9517
2.2 2.3117
};
\addplot [semithick, darkslategray54, forget plot]
table {%
2.102 0.0263
2.298 0.0263
};
\addplot [semithick, darkslategray54, forget plot]
table {%
2.102 2.3117
2.298 2.3117
};
\addplot [semithick, darkslategray54, forget plot]
table {%
-0.396 0.0393
-0.004 0.0393
};
\addplot [semithick, darkslategray54, forget plot]
table {%
0.004 0.0397
0.396 0.0397
};
\addplot [semithick, darkslategray54, forget plot]
table {%
0.604 0.0396
0.996 0.0396
};
\addplot [semithick, darkslategray54, forget plot]
table {%
1.004 0.0447
1.396 0.0447
};
\addplot [semithick, darkslategray54, forget plot]
table {%
1.604 0.0402
1.996 0.0402
};
\addplot [semithick, darkslategray54, forget plot]
table {%
2.004 0.0623
2.396 0.0623
};

\end{axis}

\draw [dashed] (c1) -- (ax2.south west);
\draw [dashed] (c2) -- (ax2.north east);

\end{tikzpicture}
        \vspace{-0.4em}
        \label{fig:box_100}
	}\\
	\subfloat[][{MS} selection policy.]
	{
\begin{tikzpicture}

\definecolor{darkgoldenrod19413532}{RGB}{194,135,32}
\definecolor{darkgray176}{RGB}{176,176,176}
\definecolor{darkslategray54}{RGB}{54,54,54}
\definecolor{lightgray204}{RGB}{204,204,204}
\definecolor{teal23108155}{RGB}{23,108,155}

\begin{axis}[
width = \columnwidth,
height = 4.5cm,
legend cell align={left},
legend style={
  fill opacity=0.8,
  draw opacity=1,
  text opacity=1,
  at={(0.03,0.97)},
  anchor=north west,
  draw=lightgray204
},
tick align=outside,
tick pos=left,
x grid style={darkgray176},
xlabel={Number of \glspl{gv} ($n$) },
xmin=-0.5, xmax=2.5,
xtick style={color=black},
xtick={0,1,2},
xticklabels={10,50,100},
y grid style={darkgray176},
ymajorgrids,
ylabel={Delay ($t_d$) [s]},
ymin=-0.1, ymax=60,
ytick style={color=black},
name=ax1
]
\path [draw=darkslategray54, fill=teal23108155, semithick]
(axis cs:-0.396,0.0364)
--(axis cs:-0.004,0.0364)
--(axis cs:-0.004,0.0411)
--(axis cs:-0.396,0.0411)
--(axis cs:-0.396,0.0364)
--cycle;
\path [draw=darkslategray54, fill=darkgoldenrod19413532, semithick]
(axis cs:0.004,0.0372)
--(axis cs:0.396,0.0372)
--(axis cs:0.396,0.0426)
--(axis cs:0.004,0.0426)
--(axis cs:0.004,0.0372)
--cycle;
\path [draw=darkslategray54, fill=teal23108155, semithick]
(axis cs:0.604,0.0369)
--(axis cs:0.996,0.0369)
--(axis cs:0.996,0.0423)
--(axis cs:0.604,0.0423)
--(axis cs:0.604,0.0369)
--cycle;
\path [draw=darkslategray54, fill=darkgoldenrod19413532, semithick]
(axis cs:1.004,0.041)
--(axis cs:1.396,0.041)
--(axis cs:1.396,0.0766)
--(axis cs:1.004,0.0766)
--(axis cs:1.004,0.041)
--cycle;
\path [draw=darkslategray54, fill=teal23108155, semithick]
(axis cs:1.604,0.0378)
--(axis cs:1.996,0.0378)
--(axis cs:1.996,0.0446)
--(axis cs:1.604,0.0446)
--(axis cs:1.604,0.0378)
--cycle;
\path [draw=darkslategray54, fill=darkgoldenrod19413532, semithick]
(axis cs:2.004,0.0475)
--(axis cs:2.396,0.0475)
--(axis cs:2.396,22.4009)
--(axis cs:2.004,22.4009)
--(axis cs:2.004,0.0475)
--cycle;
\draw[draw=darkslategray54,fill=teal23108155,line width=0.3pt] (axis cs:0,0) rectangle (axis cs:0,0);

\draw[draw=darkslategray54,fill=darkgoldenrod19413532,line width=0.3pt] (axis cs:0,0) rectangle (axis cs:0,0);

\addplot [semithick, darkslategray54, forget plot]
table {%
-0.2 0.0364
-0.2 0.0294
};
\addplot [semithick, darkslategray54, forget plot]
table {%
-0.2 0.0411
-0.2 0.0481
};
\addplot [semithick, darkslategray54, forget plot]
table {%
-0.298 0.0294
-0.102 0.0294
};
\addplot [semithick, darkslategray54, forget plot]
table {%
-0.298 0.0481
-0.102 0.0481
};
\addplot [semithick, darkslategray54, forget plot]
table {%
0.2 0.0372
0.2 0.0291
};
\addplot [semithick, darkslategray54, forget plot]
table {%
0.2 0.0426
0.2 0.0507
};
\addplot [semithick, darkslategray54, forget plot]
table {%
0.102 0.0291
0.298 0.0291
};
\addplot [semithick, darkslategray54, forget plot]
table {%
0.102 0.0507
0.298 0.0507
};
\addplot [semithick, darkslategray54, forget plot]
table {%
0.8 0.0369
0.8 0.0289
};
\addplot [semithick, darkslategray54, forget plot]
table {%
0.8 0.0423
0.8 0.0503
};
\addplot [semithick, darkslategray54, forget plot]
table {%
0.702 0.0289
0.898 0.0289
};
\addplot [semithick, darkslategray54, forget plot]
table {%
0.702 0.0503
0.898 0.0503
};
\addplot [semithick, darkslategray54, forget plot]
table {%
1.2 0.041
1.2 0.0285
};
\addplot [semithick, darkslategray54, forget plot]
table {%
1.2 0.0766
1.2 0.13
};
\addplot [semithick, darkslategray54, forget plot]
table {%
1.102 0.0285
1.298 0.0285
};
\addplot [semithick, darkslategray54, forget plot]
table {%
1.102 0.13
1.298 0.13
};
\addplot [semithick, darkslategray54, forget plot]
table {%
1.8 0.0378
1.8 0.0278
};
\addplot [semithick, darkslategray54, forget plot]
table {%
1.8 0.0446
1.8 0.0548
};
\addplot [semithick, darkslategray54, forget plot]
table {%
1.702 0.0278
1.898 0.0278
};
\addplot [semithick, darkslategray54, forget plot]
table {%
1.702 0.0548
1.898 0.0548
};
\addplot [semithick, darkslategray54, forget plot]
table {%
2.2 0.0475
2.2 0.0281
};
\addplot [semithick, darkslategray54, forget plot]
table {%
2.2 22.4009
2.2 55.9304
};
\addplot [semithick, darkslategray54, forget plot]
table {%
2.102 0.0281
2.298 0.0281
};
\addplot [semithick, darkslategray54, forget plot]
table {%
2.102 55.9304
2.298 55.9304
};
\addplot [semithick, darkslategray54, forget plot]
table {%
-0.396 0.0386
-0.004 0.0386
};
\addplot [semithick, darkslategray54, forget plot]
table {%
0.004 0.0396
0.396 0.0396
};
\addplot [semithick, darkslategray54, forget plot]
table {%
0.604 0.0394
0.996 0.0394
};
\addplot [semithick, darkslategray54, forget plot]
table {%
1.004 0.0481
1.396 0.0481
};
\addplot [semithick, darkslategray54, forget plot]
table {%
1.604 0.0406
1.996 0.0406
};
\addplot [semithick, darkslategray54, forget plot]
table {%
2.004 0.5714
2.396 0.5714
};

\coordinate (c1) at (axis cs:-0.396,0);
\coordinate (c2) at (axis cs:2.396,0.1);
\draw (c1) rectangle (axis cs:2.396,0.1);

\end{axis}

\begin{axis}[
    width = \textwidth/3.7,
    height = 2.5cm,
   name=ax2,
   scaled ticks=false,
   xmin=-0.45,xmax=2.5,
   ymin=0,ymax=0.15,
   at={($(ax1.south west)+(1cm,1.5cm)$)},
   xmajorgrids=true,
   xtick={0, 1, 2},
   xticklabels={10, 50, 100},
   y grid style={darkgray176},
   ymajorgrids,
   ytick style={color=black},
   ytick={0, 0.05, 0.1},
   yticklabels={0, 0.05, 0.1},
   clip=true
 ]
\path [draw=darkslategray54, fill=teal23108155, semithick]
(axis cs:-0.396,0.0364)
--(axis cs:-0.004,0.0364)
--(axis cs:-0.004,0.0411)
--(axis cs:-0.396,0.0411)
--(axis cs:-0.396,0.0364)
--cycle;
\path [draw=darkslategray54, fill=darkgoldenrod19413532, semithick]
(axis cs:0.004,0.0372)
--(axis cs:0.396,0.0372)
--(axis cs:0.396,0.0426)
--(axis cs:0.004,0.0426)
--(axis cs:0.004,0.0372)
--cycle;
\path [draw=darkslategray54, fill=teal23108155, semithick]
(axis cs:0.604,0.0369)
--(axis cs:0.996,0.0369)
--(axis cs:0.996,0.0423)
--(axis cs:0.604,0.0423)
--(axis cs:0.604,0.0369)
--cycle;
\path [draw=darkslategray54, fill=darkgoldenrod19413532, semithick]
(axis cs:1.004,0.041)
--(axis cs:1.396,0.041)
--(axis cs:1.396,0.0766)
--(axis cs:1.004,0.0766)
--(axis cs:1.004,0.041)
--cycle;
\path [draw=darkslategray54, fill=teal23108155, semithick]
(axis cs:1.604,0.0378)
--(axis cs:1.996,0.0378)
--(axis cs:1.996,0.0446)
--(axis cs:1.604,0.0446)
--(axis cs:1.604,0.0378)
--cycle;
\path [draw=darkslategray54, fill=darkgoldenrod19413532, semithick]
(axis cs:2.004,0.0475)
--(axis cs:2.396,0.0475)
--(axis cs:2.396,22.4009)
--(axis cs:2.004,22.4009)
--(axis cs:2.004,0.0475)
--cycle;
\draw[draw=darkslategray54,fill=teal23108155,line width=0.3pt] (axis cs:0,0) rectangle (axis cs:0,0);

\draw[draw=darkslategray54,fill=darkgoldenrod19413532,line width=0.3pt] (axis cs:0,0) rectangle (axis cs:0,0);

\addplot [semithick, darkslategray54, forget plot]
table {%
-0.2 0.0364
-0.2 0.0294
};
\addplot [semithick, darkslategray54, forget plot]
table {%
-0.2 0.0411
-0.2 0.0481
};
\addplot [semithick, darkslategray54, forget plot]
table {%
-0.298 0.0294
-0.102 0.0294
};
\addplot [semithick, darkslategray54, forget plot]
table {%
-0.298 0.0481
-0.102 0.0481
};
\addplot [semithick, darkslategray54, forget plot]
table {%
0.2 0.0372
0.2 0.0291
};
\addplot [semithick, darkslategray54, forget plot]
table {%
0.2 0.0426
0.2 0.0507
};
\addplot [semithick, darkslategray54, forget plot]
table {%
0.102 0.0291
0.298 0.0291
};
\addplot [semithick, darkslategray54, forget plot]
table {%
0.102 0.0507
0.298 0.0507
};
\addplot [semithick, darkslategray54, forget plot]
table {%
0.8 0.0369
0.8 0.0289
};
\addplot [semithick, darkslategray54, forget plot]
table {%
0.8 0.0423
0.8 0.0503
};
\addplot [semithick, darkslategray54, forget plot]
table {%
0.702 0.0289
0.898 0.0289
};
\addplot [semithick, darkslategray54, forget plot]
table {%
0.702 0.0503
0.898 0.0503
};
\addplot [semithick, darkslategray54, forget plot]
table {%
1.2 0.041
1.2 0.0285
};
\addplot [semithick, darkslategray54, forget plot]
table {%
1.2 0.0766
1.2 0.13
};
\addplot [semithick, darkslategray54, forget plot]
table {%
1.102 0.0285
1.298 0.0285
};
\addplot [semithick, darkslategray54, forget plot]
table {%
1.102 0.13
1.298 0.13
};
\addplot [semithick, darkslategray54, forget plot]
table {%
1.8 0.0378
1.8 0.0278
};
\addplot [semithick, darkslategray54, forget plot]
table {%
1.8 0.0446
1.8 0.0548
};
\addplot [semithick, darkslategray54, forget plot]
table {%
1.702 0.0278
1.898 0.0278
};
\addplot [semithick, darkslategray54, forget plot]
table {%
1.702 0.0548
1.898 0.0548
};
\addplot [semithick, darkslategray54, forget plot]
table {%
2.2 0.0475
2.2 0.0281
};
\addplot [semithick, darkslategray54, forget plot]
table {%
2.2 22.4009
2.2 55.9304
};
\addplot [semithick, darkslategray54, forget plot]
table {%
2.102 0.0281
2.298 0.0281
};
\addplot [semithick, darkslategray54, forget plot]
table {%
2.102 55.9304
2.298 55.9304
};
\addplot [semithick, darkslategray54, forget plot]
table {%
-0.396 0.0386
-0.004 0.0386
};
\addplot [semithick, darkslategray54, forget plot]
table {%
0.004 0.0396
0.396 0.0396
};
\addplot [semithick, darkslategray54, forget plot]
table {%
0.604 0.0394
0.996 0.0394
};
\addplot [semithick, darkslategray54, forget plot]
table {%
1.004 0.0481
1.396 0.0481
};
\addplot [semithick, darkslategray54, forget plot]
table {%
1.604 0.0406
1.996 0.0406
};
\addplot [semithick, darkslategray54, forget plot]
table {%
2.004 0.5714
2.396 0.5714
};

\end{axis}

\draw [dashed] (c1) -- (ax2.south west);
\draw [dashed] (c2) -- (ax2.north east);

\end{tikzpicture}
        \vspace{-0.4em}
        \label{fig:box_50}
	}
	\caption{Delay vs. $n$ for different satellite selection policies vs. $r$, with $C_{\rm LEO}=10$~TFLOPS. We consider LDBOO offloading.\vspace{-0.1cm}}
	\label{fig:boxplots}
    \vspace{-1.5em}
\end{figure}
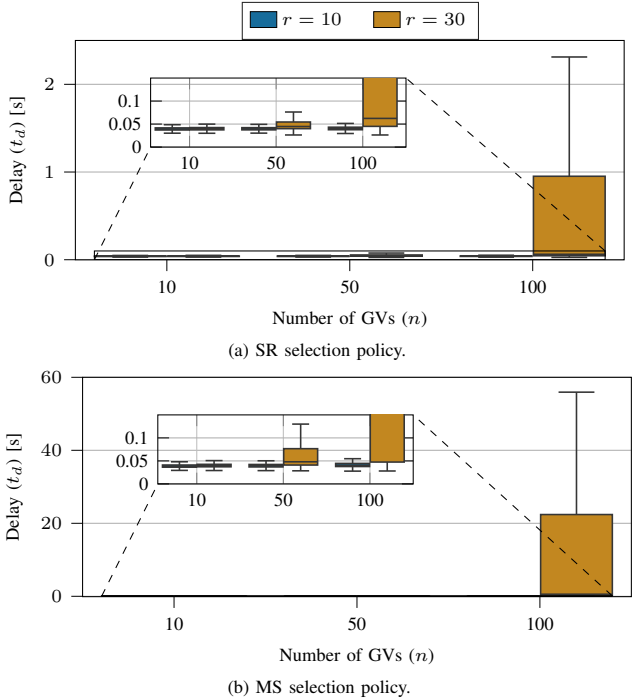

\cref{fig:elev_time} illustrates the evolution of the elevation angle $\theta$ of some Starlink satellites over time in a single simulation run, obtained from \gls{tle} data, specifically using the {SR} satellite selection policy.
As expected, we see that the visibility period of a satellite is generally short, due to the inherent mobility of the satellite in low orbits.  
Based on previous findings, good communication quality requires $\theta>50^\circ$, so the visibility is of only a few minutes. 
Therefore, satellites are required to operate in dense constellations, as in the case of Starlink, and GVs need to implement periodic handovers to maintain service continuity. 
We observe that the elevation angle changes rapidly when the satellite approaches $\theta=90^\circ$, i.e., the zenith. 

In~\cref{fig:ecdf_plots} we compare the performance of BOO vs. LDBOO by analyzing the real-time probability $P_{\rm RT}$, i.e., the probability that $t_d<\delta$, and the data drop probability $P_{\rm D}$, for different values of $t^m_o$ and $\sigma$. We set $r=30$~fps, $C_{\rm LEO} = 20$~TFLOPS, and $n = 100$ GVs.
We observe that the implementation of a light-drop policy in LDBOO improves the delay for $\sigma=4$ and 6 compared to BOO. In fact, LDBOO is designed to randomly discard data packets even before the LEO satellite queues become unstable: while this approach may increase the drop probability, it prevents the system from discarding entire bunches of packets during the back-off, which eventually improves $P_{\rm RT}$.
Notably, LDBOO may discard data when $\hat{t}_{\rm LEO}<\delta$, i.e., before congestion. In this case, $\hat{t}_{\rm LEO}/\delta<1$ so, according to Eq.~\eqref{eq:drop}, the dropping policy of LDBOO is particularly aggressive when $\sigma$ is small, which may increase the delay in the long term. In confirmation of this, we see in~\cref{fig:ecdf_plots} that BOO outperforms LDBOO when $\sigma=1$ or 2. 
Finally, we notice that the impact of  $t^m_o$ is not negligible. When $t^m_o$ is small, the system does not have time to decongest, and enters many consecutive back-off periods which would increase the delay. This behavior is similar to the Silly Window Syndrome in \gls{tcp}, 
which is incurred when the receiver (i.e., the LEO satellite) processes data slowly~\cite{rfc813}.
 On the other hand, when $t^m_o$ is large, data offloading is deactivated for a (long) back-off time, which would increase the drop probability with no improvements in terms of delay. Empirically, the best compromise was determined to be $t^m_o=10$, which achieves a good trade-off between $P_{\rm RT}$ and $P_{\rm D}$.



\begin{figure}[t!]
    \centering
    \subfloat[][$s=5662$ satellites.]
  {
\begin{tikzpicture}

\definecolor{darkgray176}{RGB}{176,176,176}
\definecolor{darkslategray38}{RGB}{38,38,38}

\begin{axis}[
width = \columnwidth/1.3,
height = 4.5cm,
colorbar,
colorbar style={ylabel={Real-time probability ($P_{RT}$)}},
colormap={mymap}{[1pt]
  rgb(0pt)=(0.705673158,0.01555616,0.150232812);
  rgb(1pt)=(0.752534934,0.157246067,0.184115123);
  rgb(2pt)=(0.795631745,0.24128379,0.220525627);
  rgb(3pt)=(0.834620542,0.312874446,0.259301199);
  rgb(4pt)=(0.869186849,0.378313092,0.300267182);
  rgb(5pt)=(0.89904617,0.439559467,0.343229596);
  rgb(6pt)=(0.923944917,0.49730856,0.387970225);
  rgb(7pt)=(0.943660866,0.551750968,0.434243684);
  rgb(8pt)=(0.958003065,0.602842431,0.481775914);
  rgb(9pt)=(0.966811177,0.650421156,0.530263762);
  rgb(10pt)=(0.969954137,0.694266682,0.579375448);
  rgb(11pt)=(0.96732803,0.734132809,0.628751763);
  rgb(12pt)=(0.958852946,0.769767752,0.678007945);
  rgb(13pt)=(0.944468518,0.800927443,0.726736146);
  rgb(14pt)=(0.924127593,0.827384882,0.774508472);
  rgb(15pt)=(0.897787179,0.848937047,0.820880546);
  rgb(16pt)=(0.865395197,0.86541021,0.865395561);
  rgb(17pt)=(0.833466556,0.860207984,0.901068838);
  rgb(18pt)=(0.798691636,0.849786142,0.931688648);
  rgb(19pt)=(0.761464949,0.834302879,0.956945269);
  rgb(20pt)=(0.722193294,0.813952739,0.976574709);
  rgb(21pt)=(0.681291281,0.788964712,0.990363227);
  rgb(22pt)=(0.639176211,0.759599947,0.998151185);
  rgb(23pt)=(0.596262162,0.726149107,0.999836203);
  rgb(24pt)=(0.552953156,0.688929332,0.995375608);
  rgb(25pt)=(0.509635204,0.648280772,0.98478814);
  rgb(26pt)=(0.46666708,0.604562568,0.968154911);
  rgb(27pt)=(0.424369608,0.558148092,0.945619588);
  rgb(28pt)=(0.38301334,0.50941904,0.917387822);
  rgb(29pt)=(0.342804478,0.458757618,0.883725899);
  rgb(30pt)=(0.30386891,0.406535296,0.84495867);
  rgb(31pt)=(0.26623388,0.353094838,0.801466763);
  rgb(32pt)=(0.2298057,0.298717966,0.753683153)
},
point meta max=1,
point meta min=0,
tick align=outside,
tick pos=left,
x grid style={darkgray176},
xlabel=\textcolor{darkslategray38}{Frame rate ($r$) [fps]},
xmin=0, xmax=2,
xtick style={color=darkslategray38},
xtick={0.5,1.5},
xticklabels={10,30},
xmajorticks=true,
y dir=reverse,
y grid style={darkgray176},
ylabel=\textcolor{darkslategray38}{LEO capacity ($c_{\rm LEO}$) [TFLOPS]},
ymin=0, ymax=4,
ytick style={color=darkslategray38},
ytick={0.5,1.5,2.5,3.5},
ymajorticks=true,
yticklabels={5,10,15,20}
]
\addplot graphics [includegraphics cmd=\pgfimage,xmin=0, xmax=2, ymin=4, ymax=0] {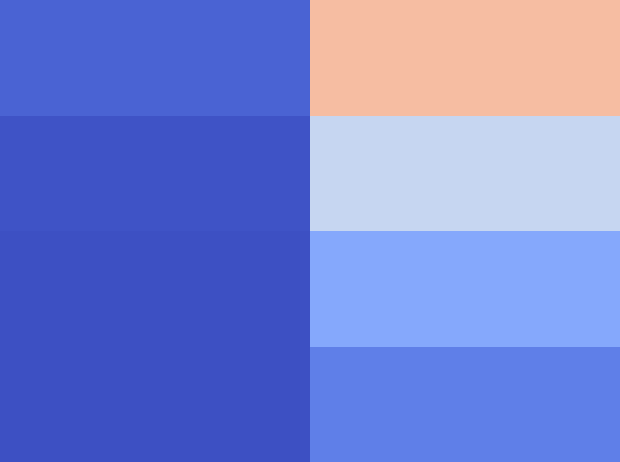};
\draw (axis cs:0.5,0.5) node[
  scale=0.5,
  text=white,
  rotate=0.0
]{0.95};
\draw (axis cs:1.5,0.5) node[
  scale=0.5,
  text=darkslategray38,
  rotate=0.0
]{0.35};
\draw (axis cs:0.5,1.5) node[
  scale=0.5,
  text=white,
  rotate=0.0
]{0.98};
\draw (axis cs:1.5,1.5) node[
  scale=0.5,
  text=darkslategray38,
  rotate=0.0
]{0.58};
\draw (axis cs:0.5,2.5) node[
  scale=0.5,
  text=white,
  rotate=0.0
]{0.99};
\draw (axis cs:1.5,2.5) node[
  scale=0.5,
  text=white,
  rotate=0.0
]{0.77};
\draw (axis cs:0.5,3.5) node[
  scale=0.5,
  text=white,
  rotate=0.0
]{0.99};
\draw (axis cs:1.5,3.5) node[
  scale=0.5,
  text=white,
  rotate=0.0
]{0.88};
\end{axis}

\end{tikzpicture}
      \vspace{-0.3em}
      \label{fig:h_100}
  }\\ \vspace{-0.5em}
   \subfloat[][$s=2831$ satellites.]
  {
\begin{tikzpicture}

\definecolor{darkgray176}{RGB}{176,176,176}
\definecolor{darkslategray38}{RGB}{38,38,38}

\begin{axis}[
width = \columnwidth/1.3,
height = 4.5cm,
colorbar,
colorbar style={ylabel={Real-time probability $(P_{RT})$}},
colormap={mymap}{[1pt]
  rgb(0pt)=(0.705673158,0.01555616,0.150232812);
  rgb(1pt)=(0.752534934,0.157246067,0.184115123);
  rgb(2pt)=(0.795631745,0.24128379,0.220525627);
  rgb(3pt)=(0.834620542,0.312874446,0.259301199);
  rgb(4pt)=(0.869186849,0.378313092,0.300267182);
  rgb(5pt)=(0.89904617,0.439559467,0.343229596);
  rgb(6pt)=(0.923944917,0.49730856,0.387970225);
  rgb(7pt)=(0.943660866,0.551750968,0.434243684);
  rgb(8pt)=(0.958003065,0.602842431,0.481775914);
  rgb(9pt)=(0.966811177,0.650421156,0.530263762);
  rgb(10pt)=(0.969954137,0.694266682,0.579375448);
  rgb(11pt)=(0.96732803,0.734132809,0.628751763);
  rgb(12pt)=(0.958852946,0.769767752,0.678007945);
  rgb(13pt)=(0.944468518,0.800927443,0.726736146);
  rgb(14pt)=(0.924127593,0.827384882,0.774508472);
  rgb(15pt)=(0.897787179,0.848937047,0.820880546);
  rgb(16pt)=(0.865395197,0.86541021,0.865395561);
  rgb(17pt)=(0.833466556,0.860207984,0.901068838);
  rgb(18pt)=(0.798691636,0.849786142,0.931688648);
  rgb(19pt)=(0.761464949,0.834302879,0.956945269);
  rgb(20pt)=(0.722193294,0.813952739,0.976574709);
  rgb(21pt)=(0.681291281,0.788964712,0.990363227);
  rgb(22pt)=(0.639176211,0.759599947,0.998151185);
  rgb(23pt)=(0.596262162,0.726149107,0.999836203);
  rgb(24pt)=(0.552953156,0.688929332,0.995375608);
  rgb(25pt)=(0.509635204,0.648280772,0.98478814);
  rgb(26pt)=(0.46666708,0.604562568,0.968154911);
  rgb(27pt)=(0.424369608,0.558148092,0.945619588);
  rgb(28pt)=(0.38301334,0.50941904,0.917387822);
  rgb(29pt)=(0.342804478,0.458757618,0.883725899);
  rgb(30pt)=(0.30386891,0.406535296,0.84495867);
  rgb(31pt)=(0.26623388,0.353094838,0.801466763);
  rgb(32pt)=(0.2298057,0.298717966,0.753683153)
},
point meta max=1,
point meta min=0,
tick align=outside,
tick pos=left,
x grid style={darkgray176},
xlabel=\textcolor{darkslategray38}{Frame rate ($r$) [fps]},
xmin=0, xmax=2,
xtick style={color=darkslategray38},
xtick={0.5,1.5},
xticklabels={10,30},
xmajorticks=true,
y dir=reverse,
y grid style={darkgray176},
ylabel=\textcolor{darkslategray38}{LEO capacity ($c_{\rm LEO}$) [TFLOPS]},
ymin=0, ymax=4,
ytick style={color=darkslategray38},
ytick={0.5,1.5,2.5,3.5},
ymajorticks=true,
yticklabels={5,10,15,20}
]
\addplot graphics [includegraphics cmd=\pgfimage,xmin=0, xmax=2, ymin=4, ymax=0] {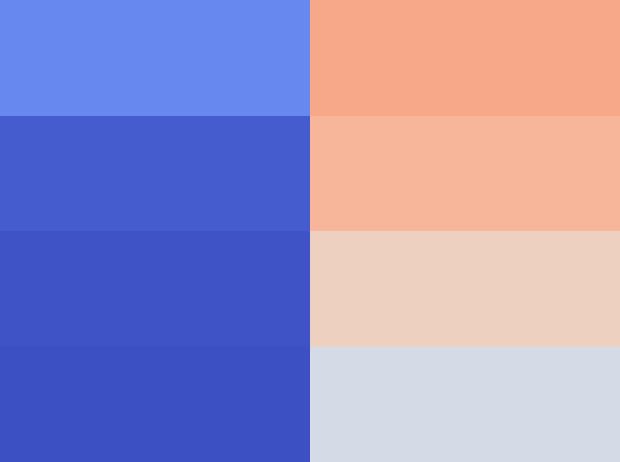};
\draw (axis cs:0.5,0.5) node[
  scale=0.5,
  text=white,
  rotate=0.0
]{0.86};
\draw (axis cs:1.5,0.5) node[
  scale=0.5,
  text=darkslategray38,
  rotate=0.0
]{0.29};
\draw (axis cs:0.5,1.5) node[
  scale=0.5,
  text=white,
  rotate=0.0
]{0.96};
\draw (axis cs:1.5,1.5) node[
  scale=0.5,
  text=darkslategray38,
  rotate=0.0
]{0.33};
\draw (axis cs:0.5,2.5) node[
  scale=0.5,
  text=white,
  rotate=0.0
]{0.98};
\draw (axis cs:1.5,2.5) node[
  scale=0.5,
  text=darkslategray38,
  rotate=0.0
]{0.42};
\draw (axis cs:0.5,3.5) node[
  scale=0.5,
  text=white,
  rotate=0.0
]{0.99};
\draw (axis cs:1.5,3.5) node[
  scale=0.5,
  text=darkslategray38,
  rotate=0.0
]{0.53};
\end{axis}

\end{tikzpicture}
    \label{fig:h_50}
    \vspace{-0.3em}
  }
    \caption{Real-time probability as a function of $r$ and $C_{\rm LEO}$, vs. the number of Starlink satellites $s$. We consider LDBOO offloading with SR.\vspace{-0.1cm}}
    \label{fig:heatmaps}
    \vspace{-1.4em}
\end{figure}
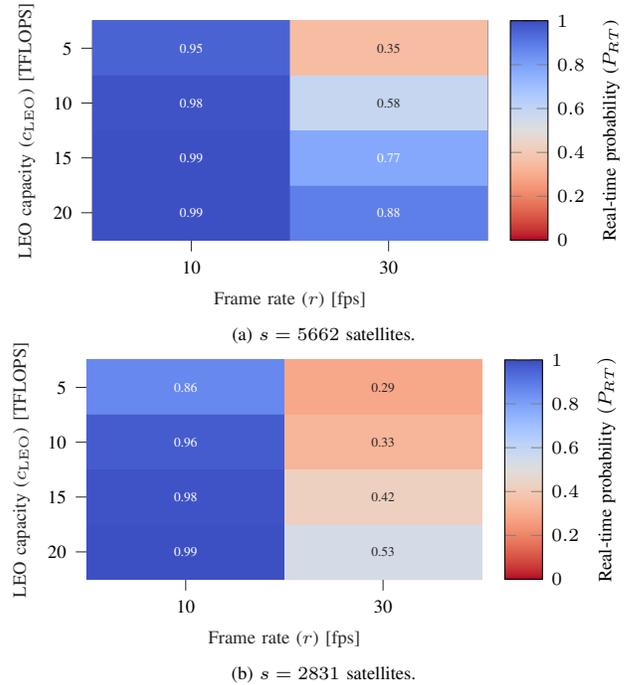

In~\cref{fig:boxplots} we plot the delay vs. $r$, for different satellite selection policies, considering LDBOO for the offloading. As expected, {SR} outperforms MS, and the gap increases especially in the high-density and/or congested scenarios. In fact, while MS tends to overload a single (i.e., the best in terms of SNR) satellite, SR tries to distribute the processing load across multiple (though possibly suboptimal) satellites, which will improve the delay in case of offloading.
The median delay is below $\delta=150$~ms in most configurations, and even below $50$~ms when $r=10$~fps.
Notice that, for $r=30$~fps and $n=100$ GVs, the system is unstable, and the delay rapidly increases up to $50$~s.
Based on the above results, we conclude that the optimal offloading strategy is LDBOO with SR, for $\sigma = 4$ and $t^m_o=10$, which will be our selected benchmark in the remainder of this paper.

In~\cref{fig:heatmaps} we investigate the impact of the size of the Starlink constellation in terms of $P_{\rm RT}$, as a function of $r$ and $C_{\rm LEO}$, for $n = 100$ GVs.
We observe that increasing the computational capacity at the LEO satellite (or, equivalently, the constellation density) is desirable, if not imperative, when $r$ increases, in order to serve processing requests in more congested scenarios, while the benefit is limited when $r=10$ fps. 

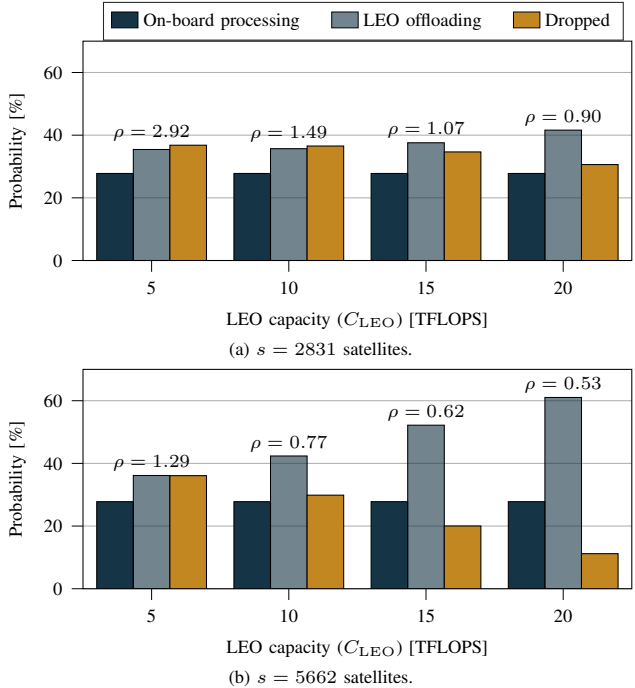
\begin{figure}[t!]
 \hspace*{14pt}
 \begin{subfigure}[b]{\columnwidth}
  \centering
%
%

\definecolor{color6giallo}{RGB}{194,135,32}
\definecolor{color1}{RGB}{21, 52, 69}

\begin{tikzpicture}
\pgfplotsset{every tick label/.append style={font=\scriptsize}}

\pgfplotsset{compat=1.11,
	/pgfplots/ybar legend/.style={
		/pgfplots/legend image code/.code={%
			\draw[##1,/tikz/.cd,yshift=-0.25em]
			(0cm,0cm) rectangle (10pt,0.6em);},
	},
}

\begin{axis}[%
width=0,
height=0,
at={(0,0)},
scale only axis,
xmin=0,
xmax=0,
xtick={},
ymin=0,
ymax=0,
ytick={},
axis background/.style={fill=white},
legend style={legend cell align=left,
              align=center,
              draw=white!15!black,
              at={(0.5, 1.3)},
              anchor=center,
              /tikz/every even column/.append style={column sep=1em}},
legend columns=3,
]
\addplot[ybar,ybar legend,draw=black,fill=color1,line width=0.08pt]
table[row sep=crcr]{%
	0	0\\
};
\addlegendentry{On-board processing}

\addplot[ybar legend,ybar,draw=black,fill=color1,fill opacity=0.6,line width=0.08pt]
  table[row sep=crcr]{%
	0	0\\
};
\addlegendentry{LEO offloading}

\addplot[ybar legend,ybar,draw=black,fill=color6giallo,line width=0.08pt]
  table[row sep=crcr]{%
	0	0\\
};
\addlegendentry{Dropped}

\end{axis}
\end{tikzpicture}%
  \end{subfigure}
    \centering
       \subfloat[][$s=2831$ satellites.]
  {
\begin{tikzpicture}

\definecolor{cadetblue105153196}{RGB}{105,153,196}
\definecolor{chocolate1869726}{RGB}{186,97,26}
\definecolor{darkgray176}{RGB}{176,176,176}
\definecolor{darkslategray66}{RGB}{66,66,66}
\definecolor{olivedrab11416083}{RGB}{114,160,83}
\definecolor{color6giallo}{RGB}{194,135,32}
\definecolor{color1}{RGB}{21, 52, 69}

\begin{axis}[
width = \columnwidth,
height = 4.5cm,
tick align=outside,
tick pos=left,
unbounded coords=jump,
x grid style={darkgray176},
xlabel={LEO capacity ($C_{\rm LEO}$) [TFLOPS]},
xmin=-0.5, xmax=3.5,
xtick style={color=black},
xtick={0,1,2,3},
xticklabels={5,10,15,20},
y grid style={darkgray176},
ylabel={Probability [\%]},
ymajorgrids,
ymin=0, ymax=70,
ytick style={color=black},
ytick={0,20,40,60,80,100},
yticklabels={0,20,40,60,80,100}
]
\draw[draw=black,fill=color1] (axis cs:-0.4,0) rectangle (axis cs:-0.133333333333333,27.7777777777778);

\draw[draw=black,fill=color1] (axis cs:0.6,0) rectangle (axis cs:0.866666666666667,27.7777777777778);
\draw[draw=black,fill=color1] (axis cs:1.6,0) rectangle (axis cs:1.86666666666667,27.7777777777778);
\draw[draw=black,fill=color1] (axis cs:2.6,0) rectangle (axis cs:2.86666666666667,27.7777777777778);
\draw[draw=black,fill=color1,fill opacity=0.6] (axis cs:-0.133333333333333,0) rectangle (axis cs:0.133333333333333,35.4291975308642);

\node[] at (axis cs:0,40.79) {\scriptsize $\rho = 2.92$};
\node[] at (axis cs:1,40.54) {\scriptsize $\rho = 1.49$};
\node[] at (axis cs:2,41.57) {\scriptsize $\rho = 1.07$};
\node[] at (axis cs:3,45.60) {\scriptsize $\rho = 0.90$};

\draw[draw=black,fill=color1,fill opacity=0.6] (axis cs:0.866666666666667,0) rectangle (axis cs:1.13333333333333,35.6857407407407);
\draw[draw=black,fill=color1,fill opacity=0.6] (axis cs:1.86666666666667,0) rectangle (axis cs:2.13333333333333,37.5730864197531);
\draw[draw=black,fill=color1,fill opacity=0.6] (axis cs:2.86666666666667,0) rectangle (axis cs:3.13333333333333,41.6042592592593);
\draw[draw=black,fill=color6giallo] (axis cs:0.133333333333333,0) rectangle (axis cs:0.4,36.793024691358);

\draw[draw=black,fill=color6giallo] (axis cs:1.13333333333333,0) rectangle (axis cs:1.4,36.5364814814815);
\draw[draw=black,fill=color6giallo] (axis cs:2.13333333333333,0) rectangle (axis cs:2.4,34.6491358024691);
\draw[draw=black,fill=color6giallo] (axis cs:3.13333333333333,0) rectangle (axis cs:3.4,30.617962962963);
\addplot [line width=1.08pt, darkslategray66]
table {%
-0.266666666666667 nan
-0.266666666666667 nan
};
\addplot [line width=1.08pt, darkslategray66]
table {%
0.733333333333333 nan
0.733333333333333 nan
};
\addplot [line width=1.08pt, darkslategray66]
table {%
1.73333333333333 nan
1.73333333333333 nan
};
\addplot [line width=1.08pt, darkslategray66]
table {%
2.73333333333333 nan
2.73333333333333 nan
};
\addplot [line width=1.08pt, darkslategray66]
table {%
0 nan
0 nan
};
\addplot [line width=1.08pt, darkslategray66]
table {%
1 nan
1 nan
};
\addplot [line width=1.08pt, darkslategray66]
table {%
2 nan
2 nan
};
\addplot [line width=1.08pt, darkslategray66]
table {%
3 nan
3 nan
};
\addplot [line width=1.08pt, darkslategray66]
table {%
0.266666666666667 nan
0.266666666666667 nan
};
\addplot [line width=1.08pt, darkslategray66]
table {%
1.26666666666667 nan
1.26666666666667 nan
};
\addplot [line width=1.08pt, darkslategray66]
table {%
2.26666666666667 nan
2.26666666666667 nan
};
\addplot [line width=1.08pt, darkslategray66]
table {%
3.26666666666667 nan
3.26666666666667 nan
};
\end{axis}

\end{tikzpicture}
    \vspace{-0.5em}
    \label{fig:bar_50}
  }\\ \vskip 0.1cm
    \subfloat[][$s=5662$ satellites.]
  {
\begin{tikzpicture}

\definecolor{cadetblue105153196}{RGB}{105,153,196}
\definecolor{chocolate1869726}{RGB}{186,97,26}
\definecolor{darkgray176}{RGB}{176,176,176}
\definecolor{darkslategray66}{RGB}{66,66,66}
\definecolor{olivedrab11416083}{RGB}{114,160,83}
\definecolor{color6giallo}{RGB}{194,135,32}
\definecolor{color1}{RGB}{21, 52, 69}

\begin{axis}[
width = \columnwidth,
height = 4.5cm,
tick align=outside,
tick pos=left,
unbounded coords=jump,
x grid style={darkgray176},
xlabel={LEO capacity ($C_{\rm LEO}$) [TFLOPS]},
xmin=-0.5, xmax=3.5,
xtick style={color=black},
xtick={0,1,2,3},
xticklabels={5,10,15,20},
y grid style={darkgray176},
ylabel={Probability [\%]},
ymajorgrids,
ymin=0, ymax=70,
ytick style={color=black},
ytick={0,20,40,60,80,100},
yticklabels={0,20,40,60,80,100}
]
\draw[draw=black,fill=color1] (axis cs:-0.4,0) rectangle (axis cs:-0.133333333333333,27.7777777777778);
\draw[draw=black,fill=color1] (axis cs:0.6,0) rectangle (axis cs:0.866666666666667,27.7777777777778);
\draw[draw=black,fill=color1] (axis cs:1.6,0) rectangle (axis cs:1.86666666666667,27.7777777777778);
\draw[draw=black,fill=color1] (axis cs:2.6,0) rectangle (axis cs:2.86666666666667,27.7777777777778);
\draw[draw=black,fill=color1,fill opacity=0.6] (axis cs:-0.133333333333333,0) rectangle (axis cs:0.133333333333333,36.1391666666667);

\node[] at (axis cs:0,40.13) {\scriptsize $\rho = 1.29$};
\node[] at (axis cs:1,46.36) {\scriptsize $\rho = 0.77$};
\node[] at (axis cs:2,56.19) {\scriptsize $\rho = 0.62$};
\node[] at (axis cs:3,65.03) {\scriptsize $\rho = 0.53$};

\draw[draw=black,fill=color1,fill opacity=0.6] (axis cs:0.866666666666667,0) rectangle (axis cs:1.13333333333333,42.361975308642);
\draw[draw=black,fill=color1,fill opacity=0.6] (axis cs:1.86666666666667,0) rectangle (axis cs:2.13333333333333,52.1907716049383);
\draw[draw=black,fill=color1,fill opacity=0.6] (axis cs:2.86666666666667,0) rectangle (axis cs:3.13333333333333,61.0349074074074);
\draw[draw=black,fill=color6giallo] (axis cs:0.133333333333333,0) rectangle (axis cs:0.4,36.0830555555556);

\draw[draw=black,fill=color6giallo] (axis cs:1.13333333333333,0) rectangle (axis cs:1.4,29.8602469135802);
\draw[draw=black,fill=color6giallo] (axis cs:2.13333333333333,0) rectangle (axis cs:2.4,20.0314506172839);
\draw[draw=black,fill=color6giallo] (axis cs:3.13333333333333,0) rectangle (axis cs:3.4,11.1873148148148);
\addplot [line width=1.08pt, darkslategray66]
table {%
-0.266666666666667 nan
-0.266666666666667 nan
};
\addplot [line width=1.08pt, darkslategray66]
table {%
0.733333333333333 nan
0.733333333333333 nan
};
\addplot [line width=1.08pt, darkslategray66]
table {%
1.73333333333333 nan
1.73333333333333 nan
};
\addplot [line width=1.08pt, darkslategray66]
table {%
2.73333333333333 nan
2.73333333333333 nan
};
\addplot [line width=1.08pt, darkslategray66]
table {%
0 nan
0 nan
};
\addplot [line width=1.08pt, darkslategray66]
table {%
1 nan
1 nan
};
\addplot [line width=1.08pt, darkslategray66]
table {%
2 nan
2 nan
};
\addplot [line width=1.08pt, darkslategray66]
table {%
3 nan
3 nan
};
\addplot [line width=1.08pt, darkslategray66]
table {%
0.266666666666667 nan
0.266666666666667 nan
};
\addplot [line width=1.08pt, darkslategray66]
table {%
1.26666666666667 nan
1.26666666666667 nan
};
\addplot [line width=1.08pt, darkslategray66]
table {%
2.26666666666667 nan
2.26666666666667 nan
};
\addplot [line width=1.08pt, darkslategray66]
table {%
3.26666666666667 nan
3.26666666666667 nan
};
\end{axis}

\end{tikzpicture}
    \vspace{-0.5em}
    \label{fig:bar_100}
  }
    \caption{Onboard processing vs. offloading vs. data drop vs. $C_{\rm LEO}$, as a function of $s$. The stability factor $\rho$ is reported on top of the bars. We consider LDBOO offloading with SR, and $r=10$ fps.\vspace{-1.8em}}
    \label{fig:barplots}
\end{figure}

Finally, \cref{fig:barplots} illustrates the probability of onboard processing vs. offloading vs. data drop, to represent how (and where) data are processed in the system.
On top of the bars, we report the load factor $\rho$ of the queues at the LEO satellites.
First, we observe that queues are often unstable, i.e., $\rho>1$, with $s = 2831$ satellites. This motivates economic investments towards dense satellite constellations, which is consistent with Starlink's future deployment plans.
Second, the probability of onboard processing does not depend on either $C_{\rm LEO}$ or $s$, but only depends on $\delta$ and $r$. In fact, regardless of the configuration of the satellite constellation, data can be processed onboard as long as the application delay requirement is satisfied, given a certain application rate.
In general, we can see that onboard processing alone can support only 30\% of traffic vs. up to around 90\% when combined with offloading.
Moreover, as $C_{\rm LEO}$ increases, more data streams can be offloaded to the satellite, and the drop probability is less than 10\%.

\vspace{-0.1cm}
\section{Conclusions and Future Works}
\label{sec:conclusions_and_future_works}
\vspace{-0.1em}
Satellite communication bridges the connectivity gap in rural and remote regions, where traditional terrestrial networks are often unavailable. In particular, \glspl{gv} can offload some computational tasks to LEO satellites in a reasonable time, a concept referred to as \gls{vec}. 
This study successfully demonstrates that Starlink satellites, if equipped with sufficient computing capacity (e.g., in terms of \glspl{gpu}), can act as space edge servers for processing ground data in real time, i.e., within the time constraints of the application. We compared several offloading schemes as a function of the application rate, the number of \glspl{gv}, and the computational resources of the system. 
We show via simulations that our proposed LDBOO scheme, which introduces controlled data dropping and periodic back-off to prevent buffer overflow at the satellites, can improve the total delay for data processing compared to onboard processing~alone. 

As part of our future work, we will design more sophisticated offloading strategies that include energy efficiency, in addition to the delay, in the optimization.

\bibliographystyle{IEEEtran}
\bibliography{bibliography.bib}

\end{document}